# Flood propagation modelling with the Local Inertia Approximation: theoretical and numerical analysis of its physical limitations

Luca Cozzolino[1], Luigi Cimorelli[2], Renata Della Morte[3], Giovanni Pugliano[4], Vincenzo Piscopo[5], Domenico Pianese[6]

**Abstract**

*Attention of the researchers has increased towards a simplification of the complete Shallow water Equations called the Local Inertia Approximation (LInA), which is obtained by neglecting the advection term in the momentum conservation equation. This model, whose physical basis is discussed here, is commonly used for the simulation of slow flooding phenomena characterized by small velocities and absence of flow discontinuities. In the present paper it is demonstrated that a shock is always developed at moving wetting-drying frontiers, and this justifies the study of the*

[1] Sen. Res., Ph. D., Dept. of Engrg., Parthenope Univ., Centro Direzionale di Napoli – Is. C4, 80143 Napoli, Italy. E-mail: luca.cozzolino@uniparthenope.it
[2] Sen. Res., Ph. D., Dept. of Civil, Architectural and Environmental Engrg., Federico II Univ., via Claudio 21, 80125 Napoli, Italy. E-mail: luigi.cimorelli@unina.it
[3] Full Prof., Ph. D., Dept. of Engrg., Parthenope Univ., Centro Direzionale di Napoli – Is. C4, 80143 Napoli, Italy. E-mail: renata.dellamorte@uniparthenope.it
[4] Ass. Prof., Ph. D., Dept. of Engrg., Parthenope Univ., Centro Direzionale di Napoli – Is. C4, 80143 Napoli, Italy. E-mail: giovanni.pugliano@uniparthenope.it
[5] Jun. Res., Ph. D., Dept. of Sci. and Tech., Parthenope Univ., Centro Direzionale di Napoli – Is. C4, 80143 Napoli, Italy. E-mail: vincenzo.piscopo@uniparthenope.it
[6] Full Prof., Dept. of Civil, Architectural and Environmental Engrg., Federico II Univ., via Claudio 21, 80125 Napoli, Italy. E-mail: pianese@unina.it

*Riemann problem on even and uneven beds. In particular, the general exact solution for the Riemann problem on horizontal frictionless bed is given, together with the exact solution of the non-breaking wave propagating on horizontal bed with friction, while some example solution is given for the Riemann problem on discontinuous bed. From this analysis, it follows that drying of the wet bed is forbidden in the LInA model, and that there are initial conditions for which the Riemann problem has no solution on smoothly varying bed. In addition, propagation of the flood on discontinuous sloping bed is impossible if the bed drops height have the same order of magnitude of the moving-frontier shock height. Finally, it is found that the conservation of the mechanical energy is violated. It is evident that all these findings pose a severe limit to the application of the model. The numerical analysis has confirmed the existence of the frontal shock in advancing flows, but has also demonstrated that LInA numerical models may produce numerical solutions, which are unreliable because of mere algorithmic nature, also in the case that the LInA mathematical solutions do not exist.*

*Following the preceding results, two criteria for the definition of the applicability limits of the LInA model have been considered. These criteria, which are valid for the very restrictive case of continuously varying bed elevation, are based on the limitation of the wetting front velocity and the limitation of spurious total head variations, respectively. Based on these criteria, the applicability limits of the LInA model are discouragingly severe, even if the bed elevation varies continuously. More important, the non-existence of the LInA solution in the case of discontinuous topography and the non-existence of receding fronts radically question the viability of the LInA model in realistic cases. It is evident that classic SWE models should be preferred in the majority of the practical applications.*

Corresponding author: Luca Cozzolino

E-mail address: luca.cozzolino@uniparthenope.it

Address: Dipartimento di Ingegneria, Università degli Studi di Napoli Parthenope, Isola C4, 80143 Napoli (Italy)

Phone: +390815476723.

# 1 Introduction

It is well recognized that flooding risk is now a global issue, with more than $2.8 \cdot 10^9$ people affected in the years from 1980 to 2009 (Doocy et al. 2013). There is statistical evidence that the climate change is driving the increase of heavy precipitation events (Kundzewich et al. 2014), leading to an increase of the flooding hazard. On the other side, flood risk is becoming an issue in urban areas, due to the rapid growing of population in the cities and the steady anthropic pressure on the environment (Chen et al. 2015). These observations highlight the need of implementing robust and physically based models for the evaluation of flooded area extents, and for the evaluation of countermeasures efficiency.

The Shallow water Equations (SWE) are the elective mathematical model for the computation of flood propagation in floodplains (Liang et a. 2008), urban flooding (Mignot et al. 2006), tsunami propagation in coastal areas (LeVeque et al. 2011), and dam-break computation (Wang et al. 2011). This versatility of use is justified by the ability of the SWE to cope with real-flow physical features such as supercritical, subcritical, as well as transcritical flow conditions. In addition, the SWE can manage with moving flow field discontinuities (Montuori and Greco 1973), discontinuous bed elevation (Alcrudo and Benkhaldoun 2001, Cozzolino et al. 2017), and wetting-drying fronts (Sobey 2009). The two dimensional SWE satisfy the rotational invariance property (Toro 2001), and for this reason the one-dimensional SWE model

$$\begin{aligned}
&\frac{\partial h}{\partial t} + \frac{\partial hu}{\partial x} = 0 \\
&\frac{\partial hu}{\partial t} + \frac{\partial}{\partial x}\left(\frac{gh^2}{2} + hu^2\right) = -gh\frac{\partial z}{\partial x} - ghS_f
\end{aligned} \quad (1)$$

is sufficient to study the essential characteristics of the mathematical model. In Eq. (1), the symbols have the following meaning: $x$ and $t$ are the space and time independent variables, respectively; $h(x$,

$t$) is the flow depth, while $u(x, t)$ is the vertically averaged flow velocity; $g = 9.81$ m/s$^2$ is the gravity acceleration; $z(x)$ is the bed elevation; and $S_f = S_f(x, h, u)$ is the friction slope. The first of Eq. (1) represents the differential form of the mass conservation principle, while the second equation represents the second principle of the dynamics (often called momentum conservation equation). The solutions of Eq. (1) must be characterised by the flow depth-positivity property, namely the condition $h > 0$ must be satisfied.

Computation of flooding is based on a delicate equilibrium between the need of a complete physical representation and the need for fast computation, and there are two main paths followed by researchers in order to simultaneously fulfil these two competing objectives. The first path consists in the use of increasingly powerful computer architectures and parallel computing (Lacasta et al. 2015), while the second path is based on the use of simplified models derived from the SWE model (Neal et al. 2012). Recently, attention of the researchers has increased towards a simplification of the complete SWE called the Local Inertia Approximation (LInA). This system of equations, whose one-dimensional version is

$$(2) \quad \begin{aligned} &\frac{\partial h}{\partial t} + \frac{\partial hu}{\partial x} = 0 \\ &\frac{\partial hu}{\partial t} + \frac{\partial}{\partial x}\frac{gh^2}{2} = -gh\frac{\partial z}{\partial x} - ghS_f \end{aligned},$$

is obtained by neglecting the advection term in the momentum conservation equation of the SWE, while the local inertial term is retained. The LInA is a non-linear system of hyperbolic differential equations (Natale and Savi 1991) that has been first used in oceanography (Uusitalo 1960, Sielecki 1963) for storm-surge computation in sea basins with fixed wet-dry boundaries. Successively, its use has been extended to border irrigation computation (Natale and Savi 1991, Galbiati and Savi 1995), and wave propagation on floodplains (Aronica et al. 1998, Moramarco et al. 2005). The system of Eq. (2) has been recently used in a variety of applications, at urban (Neal et al. 2011, Martins et al.

2016c, Martins et al. 2017), basin (Coulthard et al. 2013, Cea and Bladé 2015, Savage et al. 2016, Nguyen et al. 2016), and regional scale (Falter et al. 2013, Yamazaki et al. 2013, Mateo et al. 2017).

The theoretical study of the LInA model has begun with the work by Bates et al. (2010). Neal et al. (2012) reported that the lack of the advection term in the momentum equation prevented the model from correctly representing the regions of the flow field characterized by critical or transcritical flow. de Almeida and Bates (2013) showed with steady flow numerical experiments that the LInA seemed a sufficient approximation of the SWE for subcritical flows with Froude number less than 0.5, and with Froude numbers in the range [0.5, 1] and mild flow depth gradients. As expected, they found numerically that the LInA systematically underestimates the wave front speed in flooding experiments. Martins et al. (2016b) supplied the exact solution of the dam-break for the LInA, and compared it with the corresponding solution of the SWE, showing that the LInA fails in reproducing supercritical flows. Cea and Bladé (2015) observed that the LInA model may be inaccurate when complicate geometries such as street junctions are present, and that the calibrated values of roughness in urban areas for the LInA model might be very different from those calibrated for the complete SWE. In addition, they observed that the LInA model underestimates the extension of the flooded areas with Froude number greater than about 0.8. As a result of the theoretical analyses available in the literature (Neal et al. 2012, de Almeida and Bates 2013), it is commonly believed that the LInA model "should only be applied in sub-critical flow conditions and with gradually varying flow" (Coulthard et al. 2013). Of course, the wet-dry fronts are always supercritical, due to the vanishing depth combined with finite front velocity, and this seems to pose a severe limit to the applicability of the LInA model to flooding applications. This observation justifies the examination of the LInA model, with special reference to the modelling of wet-dry frontiers, contained in the present work.

The theoretical study of the LInA model has been accompanied by the development of numerical models for the approximation of its solution. Bates et al. (2010) demonstrated that a simple raster-based explicit LInA numerical model exhibited a reduced computational burden with respect to a similar numerical implementation of the diffusive wave (Hunter et al. 2005). This result was successively confirmed by Dottori and Todini (2011) and Neal et al. (2011). The maximum time step was evaluated by Bates et al. (2010) on the basis of linear stability considerations, but no theoretical justification was given. In order to improve the stability of the model, de Almeida et al. (2012) proposed a numerical scheme (termed *q-centred* scheme) with semi-implicit treatment of the friction source term and discretisation of the momentum equation inspired by the Lax-Friedrichs model. Cea and Bladé (2015) proposed a Finite Volume model on unstructured triangular grid where bed slope and flow depth gradient terms were merged, and a simple upwinding technique was used for the calculation of interface fluxes. Martins et al. (2015) presented a Finite Volume scheme equipped with the approximate Roe solver for the computation of fluxes between cells. In the same paper, it was given a novel stability condition where both flow depth and velocity were taken into account, but again no theoretical justification was given. In Martins et al. (2016a), a Finite Volume scheme based on the Roe Riemann solver and a MacCormack finite difference scheme were tested using the dam-break exact solution. The results showed that numerical algorithms written in conservative form are able to reproduce moving discontinuities arising in LInA, but the treatment of the wetting-drying fronts was not clearly specified. Finally, numerous empirical methods for the treatment of wetting-drying problems were compared in Martins et al. (2018), but these methods are numerical procedures that are not based on the rigorous definition of a theoretical maximum time step for the satisfaction of the depth-positivity requirement.

The application of the LInA model in realistic two-dimensional problems has shown some recurring issue. Neal et al. (2012) reported that an explicit raster-based implementation of the LInA exhibited significant total conserved mass error, together with numerical instabilities at the wetting front when the values of the bed roughness were low. de Almeida et al. (2012) proved that their *q-centred* scheme could reduce numerical instabilities and improve the mass conservation property, but again some residual mass unbalance remained during a two hours realistic urban flooding simulation. Martins et al. (2017) found that a Roe Finite Volume scheme was characterized by significant mass conservation errors in a valley flooding test, due to the presence of drying cells at receding fronts, while negligible mass errors were present in the numerical tests where flooding occurred on planar inclines. In the present paper, it will be shown that these numerical issues are not connected to the specific algorithm implementation, because they descend from the mathematical model itself to be solved.

The paper is organized as follows. In the second Section, the SWE are rewritten in dimensionless form, in order to examine the physical justifiability of the LInA model, showing that the LInA model is not more accurate than the diffusive wave (called Noninertia Approximation in the following) and it is even worse at the wetting front, where non-physical multiple solutions are produced. In addition, it is shown that the free surface profile can be continuously connected to the dry bed in the case of wet-dry frontier at rest only. An interesting consequence of this result is that moving frontiers between wet and dry bed must be shocks that have no physical counterpart. In the third Section, the complete solution of the Riemann problem is supplied, with special attention to the Riemann problem on dry bed. The characterization of the moving discontinuities on the dry bed allows finding a second surprising result, namely that the LInA model supports only wetting phenomena, while authentic drying phenomena are forbidden. In other words, the LInA model does not admit receding wet-dry frontiers. In the fourth Section, the exact solution of the non-breaking wave on horizontal bed with friction is given, showing that the flooding wave is led by a shock that satisfies the theoretical treatment of the preceding section. Irregularities of the bed elevation and obstacles are commonly present in the flooded areas, and the interaction of moving flow discontinuities with these geometrical singularities is tackled in the context of the Riemann problem with discontinuous bed elevation in the fifth Section. In particular, it is shown that the solution of the LInA model does not exist if the bed drop is about 60% higher than the height of the advancing frontal shock. It is evident that the non-existence of the LInA solution in the case of discontinuous topography and the non-existence of receding fronts radically question the viability of the LInA model in realistic cases. In the sixth Section, the results of the numerical scheme by de Almeida et al. (2012) and of a novel Rusanov Finite Volume scheme are compared, showing that the numerical inconsistencies commonly discussed in the literature, such as the lack of mass conservation, are not due to the algorithms themselves, but are a direct consequence of the mathematical model adopted. All these theoretical and numerical results are discussed in the seventh Section, where it is shown that the applicability limits of the LInA model are discouragingly severe, even if the bed elevation varies continuously. In conclusion, the writers suggest that classic SWE models should be preferred in the majority of the practical applications.

## 2 Preliminary study of the LInA model

In this section, the assumptions at the basis of the LInA model are critically reviewed, and the problem of the wetting-drying frontier is introduced, highlighting fatal incongruences that are evidenced for the first time in the literature.

### *2.1 Analysis of the assumption of negligible momentum flux gradient*

A closer examination of the LInA model can be carried out by rewriting the SWE of Eq. (1) in dimensionless form with appropriate scaling parameters (Tsai 2003, Hunter et al. 2007, Fowler 2011). If the bed-slope $S_0 = -\partial z/\partial x$ is constant and the Chézy formula $S_f = u|u|/(C^2 h)$ is used for the calculation of the friction slope, the Eq. (1) can be rewritten as

$$(3) \quad \begin{aligned} & \frac{\partial H}{\partial T} + \frac{\partial HU}{\partial X} = 0 \\ & F_r^2 \left( \frac{\partial HU}{\partial T} + \frac{\partial HU^2}{\partial X} \right) + \frac{1}{2}\frac{\partial H^2}{\partial X} = H - U|U| \end{aligned},$$

where the following positions have been made:

(4) $X = x/L_r, \quad T = tu_r/L_r, \quad H = h/h_r, \quad U = u/u_r$.

In Eq. (4), $h_r$ is the normal flow depth corresponding to the reference flow velocity $u_r$ and to the bed slope $S_0$, while $L_r = h_r/S_0$ is a reference distance and $F_r = u_r/\sqrt{gh_r}$ is the reference Froude number. From Eq. (3), it is evident that both the local inertia term $\partial hu/\partial t$ and the momentum flux gradient $\partial hu^2/\partial x$ scale with the square of $F_r$, and vanish for small $F_r$.

Numerical computations in typical river flow conditions (Cunge et al. 1980, Price 1994), prove that active forces (gravity, pressure, and friction), are much greater than inertial forces (local inertia and momentum flux gradient). For this reason, inertial terms may be neglected in the small Froude number limit, obtaining the Noninertia Approximation (NIA). In most cases, the order of magnitude of local inertia is smaller than that of momentum flux gradient (Dingman 2009), and the approximate model where the local inertial term is neglected is called Quasi-Steady Dynamic Wave (Tsai 2003). These observations suggest that neglecting the momentum flux gradient only, which is of the same magnitude of the local inertia, or even greater, is not a well-justified procedure, even in the small Froude number limit.

For typical floodplain propagation conditions, one could ask which the relative importance of the forces acting on the flow is. It is evident that the observations made for rivers are valid for the main body of the flood, where the Froude number is small, and it remains to be seen what happens at the head of the flooding wave, where the Froude number is not negligible. The laboratory experiments show that the flow velocity is uniform in the tip-region of the flooding wave (Dressler 1952), and this assumption serves as the basis of the classic mathematical treatment of fast transients as the dam-break with friction (Whitham 1955). For this reason, the exact solution for floodplain wetting with friction is considered here, imposing that the flow velocity $u = u_r > 0$ is uniform in space and constant in time. Under this assumption, the Eq. (3) can be rewritten as

$$\frac{\partial H}{\partial T}+\frac{\partial H}{\partial X}=0$$

(5)
$$F_r^2\left(\frac{\partial H}{\partial T}+\frac{\partial H}{\partial X}\right)+\frac{1}{2}\frac{\partial H^2}{\partial X}=H-1,$$

because $U = 1$. Finally, after the substitution of the first of Eq. (5) into the second, the system reduces to

(6)
$$\frac{\partial H}{\partial T}+\frac{\partial H}{\partial X}=0$$
$$\frac{1}{2}\frac{\partial H^2}{\partial X}=H-1,$$

which clearly coincides with the NIA model. The first of Eq. (6) is the linear advection equation (LeVeque 1992), and it has exact solution

(7) $H(X,T)=\eta(X-T),$

where $\eta(X) = H(X,0)$ is the initial condition for the flow depth. From Eq. (7), it is immediate to see that $\partial H/\partial X = d\eta/dX$, and the substitution of this derivative into the second of Eq. (6) supplies the ordinary differential equation

(8) $\eta\dfrac{d\eta}{dX}-\eta+1=0,$

which has solution $\eta+\ln(1-\eta)=X$ when the boundary condition $\eta(0)=0$ is assumed. This solution is represented in Figure 1, where it is possible to observe how the flow depth is continuously connected to the dry bed. Notably, the flow depth tends to zero for $X \to 0^-$, while the corresponding gradient $d\eta/dX = (\eta-1)/\eta$ tends to infinity, and this means that both the friction term and the gradient of the pressure forces tend to infinity for $X \to 0^-$.

Recalling that and $\partial H/\partial T = -d\eta/dX$, the application of the same procedure to the LInA model of Eq. (2) leads to the ordinary differential equation

$$\eta \frac{d\eta}{dX} - \eta + 1 = F_r^2 \frac{d\eta}{dX}, \quad (9)$$

which has solution $\eta + (1-F_r^2)\ln(1-\eta) = X$ when the boundary condition $\eta(0) = 0$ is assumed. The corresponding LINA free surface profile, which is represented in Figure 1 for $F_r = 0.5$, is characterized by multiple values of the dimensionless flow depth $\eta$ for $X \geq 0$. The comparison between Eq. (8) and Eq. (9) shows that the inadmissible multivalued solution is generated by the presence of the local inertia term $F_r^2 d\eta/dX$, because the gradient $d\eta/dX = (\eta-1)/(\eta-F_r^2)$ of $\eta$ is positive and tends to $1/F_r^2$ for $\eta \to 0$ (see Figure 1).

In conclusion, the construction of the exact solution for the wave propagating with velocity $u = u_r > 0$ over an inclined dry bed with uniform friction shows that the SWE model reduces to the NIA model, and that the corresponding free surface profile is a single-valued function that is smoothly connected to the dry bed. The mutual cancellation of the inertial terms follows directly from the assumption of uniform and constant velocity, but the observation that the friction term and the gradient of the pressure forces tend to infinity suggests that the inertial terms at the wetting fronts are negligible with respect to the pressure and friction terms. Actually, it is possible to demonstrate that the NIA model is an asymptotic form of the SWE model at wetting fronts that move with finite velocity (Duran et al. 2015), despite the fact that the Froude number tends to infinity for vanishing depth. This mathematical result is in agreement with the experimental observations (Chanson 2005).

By contrast, the LInA model exhibits a multivalued solution under the assumption $u = u_r > 0$, and this fatal incongruence can be explained by observing that only the local inertia term is retained. This suggests that the LInA model, which is not better than the NIA model far from the wetting fronts, is even worse at the head of the flooding wave. In the theory of hyperbolic non-linear waves, the tendency to produce multivalued solutions is connected to the incipient formation of shocks (Whitham 1974), and this prompts a deeper study of wetting front dynamics in the LInA model.

[Insert Figure 1 about here]

*2.2 Wetting-drying frontier in the LInA model*

The study of the wetting-drying frontier can be conducted in the context of the classic method of characteristics for hyperbolic systems of differential equations (Abbott 1966). The trajectory of this front is $x_F = x_F(t)$, and the corresponding velocity $dx_F(t)/dt = u(x_F(t),t)$ coincides with the local flow velocity (Sobey 2009), implying that $h(x_F(t),t) = 0$ in the case that the flow is continuously connected to the dry bed. If discontinuities are absent, some working shows that Eq. (2) can be rewritten in the following characteristic form (compare with the corresponding expression in Martins et al. 2016a,b)

$$(10) \quad \begin{aligned} \frac{\partial I_1(\mathbf{u})}{\partial t} + \lambda_1(\mathbf{u})\frac{\partial I_1(\mathbf{u})}{\partial x} &= -gh\frac{\partial z}{\partial x} - ghS_f \\ \frac{\partial I_2(\mathbf{u})}{\partial t} + \lambda_2(\mathbf{u})\frac{\partial I_2(\mathbf{u})}{\partial x} &= -gh\frac{\partial z}{\partial x} - ghS_f \end{aligned},$$

where $\mathbf{u}(x,t) = (h \quad hu)^T$ is the vector of the conserved variables, $T$ is the symbol of matrix transpose, $\lambda_1(\mathbf{u})$ and $\lambda_2(\mathbf{u})$ are defined as

(11) $\lambda_1(\mathbf{u}) = -\sqrt{gh}, \quad \lambda_2(\mathbf{u}) = \sqrt{gh},$

while $I_1(\mathbf{u})$ and $I_2(\mathbf{u})$ are defined as

(12) $I_1(\mathbf{u}) = hu - \dfrac{2}{3}h\sqrt{gh}, \quad I_2(\mathbf{u}) = hu + \dfrac{2}{3}h\sqrt{gh}.$

It is possible to demonstrate the following

**Proposition 1**. *In the LInA model, the free surface profile can be continuously connected to the dry bed only in the case of wet-dry frontier at rest.*

*Proof.* The Eq. (10) can be interpreted in the sense that the information about the flow is transported by $I_1$ and $I_2$ along the characteristic curves of the plane $(x, t)$ defined by $dx/dt = \lambda_1$ and $dx/dt = \lambda_2$, respectively. The celerities $\lambda_1$ and $\lambda_2$ are null at the wetting-drying frontier with $h \to 0$, and this implies that the information originating from $x_F$ cannot propagate. In other words, the wet-dry frontier with null flow-depth cannot move.

By exclusion, an immediate consequence of Proposition 1 is a surprising result that is formulated here for the first time, namely that

**Corollary 1**. *The moving frontiers compatible with the LInA model can be shocks only.*

Interestingly, the feature of the LInA model highlighted by the *Corollary* 1 has not a physical counterpart, and it is totally in contrast with the SWE model, where the flow profile is always smoothly connected to the dry bed. In many LInA numerical simulations presented in the literature

(for example, Figures 9 and 10 in de Almeida and Bates 2013) the presence of the frontal shock is evident, but this feature has never been recognized and discussed before.

**3 Solution of the Riemann problem on horizontal bed without friction**

The *Corollary* 1 highlights the importance of studying moving discontinuities in the LInA model, and this can be accurately accomplished in the context of the Riemann problem solution. This problem is discussed in the present section, and its results are compared with those supplied by the SWE (whose complete solution can be found in Toro 2001). Special attention is given to the Riemann problem on dry bed.

*3.1 Position of the Riemann problem and general solution*

If the bed is horizontal and frictionless, the Eq. (2) can be rewritten as

$$(13) \quad \frac{\partial \mathbf{u}}{\partial t} + \frac{\partial \mathbf{f}(\mathbf{u})}{\partial x} = 0,$$

where $\mathbf{f}(\mathbf{u}) = \begin{pmatrix} hu & 0.5gh^2 \end{pmatrix}^T$ is the vector of the fluxes. The eigenvalues $\lambda_1(\mathbf{u})$ and $\lambda_2(\mathbf{u})$ of the Jacobian matrix $\mathbf{A}(\mathbf{u}) = \partial \mathbf{f}/\partial \mathbf{u}$ are defined by Eq. (11), with corresponding eigenvectors (de Almeida et al. 2012, Yamazaki et al. 2015)

$$(14) \quad \mathbf{r}_1(\mathbf{u}) = \begin{pmatrix} 1 & -\sqrt{gh} \end{pmatrix}^T, \quad \mathbf{r}_2(\mathbf{u}) = \begin{pmatrix} 1 & \sqrt{gh} \end{pmatrix}^T.$$

The eigenvectors $\mathbf{r}_1(\mathbf{u})$ and $\mathbf{r}_2(\mathbf{u})$ are real valued and linearly independent for each $h > 0$, and this implies that the system of Eq. (13) is strictly hyperbolic. It is easy to see (Appendix A) that the first and the second characteristic field are genuinely non-linear, and that they may contain either rarefactions or shock waves.

The Riemann problem consists in solving the system of Eq. (13) with the following discontinuous initial conditions:

$$(15) \quad \mathbf{u}(x,0) = \begin{cases} x < 0, & \mathbf{u}_L = (h_L \quad h_L u_L)^T \\ x > 0, & \mathbf{u}_R = (h_R \quad h_R u_R)^T \end{cases}.$$

The solution of the Riemann problem is self-similar, and then a vector function $\mathbf{v}(s)$ of the scalar parameter $s$ exists such as $\mathbf{u}(x,t) = \mathbf{v}(x/t)$. For $t > 0$, the initial left state $\mathbf{u}_L$ and the initial right state $\mathbf{u}_R$ are separated by the intermediate state $\mathbf{u}_M = (h_M \quad h_M u_M)^T$, which is in turn connected to $\mathbf{u}_L$ by means of a wave (a shock or a rarefaction) contained in the first characteristic field, and it is connected to $\mathbf{u}_R$ by a wave (a shock or a rarefaction) contained into the second characteristic field.

*3.1.1 Elementary waves*

In the plane ($h$, $hu$), the direct 1-wave curve

$$(16) \quad H_1(\mathbf{u}, \mathbf{u}_0) = \begin{cases} R_1(\mathbf{u}, \mathbf{u}_0): & 0 < h \leq h_0, \quad hu = h_0 u_0 + 2\left(h_0 \sqrt{gh_0} - h\sqrt{gh}\right)/3 \\ S_1(\mathbf{u}, \mathbf{u}_0): & h \geq h_0, \quad hu = h_0 u_0 + (h - h_0)\sigma_1(\mathbf{u}, \mathbf{u}_0) \end{cases}.$$

is defined as the locus of the right states $\mathbf{u}$ that are connected to the left state $\mathbf{u}_0$ by means of a direct rarefaction $R_1$ (Appendix B) or a direct shock $S_1$ (Appendix C) contained in the first characteristic field. It is possible to demonstrate (Appendix D) that the curve $H_1$ is continuous, strictly decreasing, and strictly concave. In a similar manner, the backward 2-wave curve

$$(17)\ H_2^B(\mathbf{u},\mathbf{u}_0) = \begin{cases} R_2^B(\mathbf{u},\mathbf{u}_0): & 0 < h \leq h_0, \quad hu = h_0 u_0 - 2\left(h_0\sqrt{gh_0} - h\sqrt{gh}\right)/3 \\ S_2^B(\mathbf{u},\mathbf{u}_0): & h > h_0, \quad hu = h_0 u_0 + (h - h_0)\sigma_2(\mathbf{u},\mathbf{u}_0) \end{cases}.$$

is defined as the locus of the left states $\mathbf{u}$ that are connected to the right state $\mathbf{u}_0$ by means of a backward rarefaction $R_2^B$ (Appendix B) or a backward shock $S_2^B$ (Appendix C) contained in the second characteristic field. The curve $H_2^B$ is continuous, strictly increasing, and strictly convex (Appendix D).

In Eqs. (16) and (17), the functions

$$(18)\ \sigma_1(\mathbf{u},\mathbf{u}_0) = -\sqrt{0.5g(h+h_0)}, \quad \sigma_2(\mathbf{u},\mathbf{u}_0) = \sqrt{0.5g(h+h_0)}$$

are the speeds of the shocks contained into the first and the second characteristic field (Appendix C), respectively. It is immediate to show the following

**Proposition 2**. *In the LInA model, the speed of the shocks contained into the first [second] characteristic field of the LInA is always negative [positive].*

*Proof.* The proposition follows from the signs in Eq. (18).

This is very different from what happens in the SWE, where the sign of the shock speeds depends on the flow characteristics on the two sides of the discontinuity.

*3.1.2 Dry bed state and cavitation*

In the SWE, the Riemann problem admits solutions where the dry bed is formed also if the bed is initially wet everywhere (Toro 2001). It is interesting to verify if this type of solution, called cavitation, is admissible for the Riemann problem in the LInA equations.

From Eqs. (16) and (17) it is possible to observe that the generic state can be connected to the dry bed state $\mathbf{0} = (0 \quad 0)^T$ by means of a shock, but not by a rarefaction. In particular, the inspection of Eq. (16) shows that the curve of the right states $\mathbf{u}$ connected to the left dry bed state $\mathbf{0}$ coincides with the direct shock curve

(19) $S_1(\mathbf{u}, \mathbf{0}): \quad h \geq 0, \quad hu = h\sigma_1(\mathbf{u}, \mathbf{0})$.

Similarly, the inspection of the Eq. (17) shows that the curve of the left states $\mathbf{u}$ connected to the right dry bed state $\mathbf{0} = (0 \quad 0)^T$ coincides with the backward shock curve

(20) $S_2^B(\mathbf{u}, \mathbf{0}): \quad h \geq 0, \quad hu = h\sigma_2(\mathbf{u}, \mathbf{0})$.

The flow velocity corresponding to the state $\mathbf{u}$ behind a shock that advances on the dry bed coincides with the shock celerity, i.e. $u = (hu)/h = \sigma_i(\mathbf{u}, \mathbf{0})$ with $i = 1, 2$. In addition, the Froude number $F(\mathbf{u}) = u/\sqrt{gh}$ corresponding to the state $\mathbf{u}$ connected to the dry bed is constant along the curves $S_1(\mathbf{u}, \mathbf{0})$ and $S_2^B(\mathbf{u}, \mathbf{0})$, and its absolute value is $F_w = \sqrt{0.5} \approx 0.707$.

The presence of a shock connected to the dry bed is opposite to what happens in the SWE, where a generic state is always connected to the dry bed by means of a rarefaction. An interesting result about the direction of the shock on dry bed is the following

**Proposition 3**. *In the LInA model, only shocks advancing on the dry bed are admissible, while receding shocks are forbidden.*

*Proof.* The left dry bed state **0** is connected to the right state **u** by a shock contained into the first characteristic field (see Eq. [19]), which has negative celerity (see *Proposition 2*). It follows that a shock on dry bed contained into the first characteristic field must always advance. A symmetric proof is valid for the shocks on dry bed contained into the second characteristic field.

An immediate consequence is the following conclusion:

**Proposition 4**. *In the LInA model, the drying of the wet bed is forbidden.*

*Proof.* The proposition is a consequence of *Corollary 1* and *Proposition 3*.

Notably, the *Proposition 4* is totally general and it is not a consequence of special initial conditions. In particular, the properties of the shocks defined by Eqs. (18)-(20) are valid also in the case of sloping bed with friction, because these effects are negligible through the small length of the discontinuity. This means that the LInA model cannot support an authentic bed-drying phenomenon by means of a moving wet-dry frontier. This is in contrast with the SWE model, where the drying of the wet bed is possible.

*3.1.3 General solution of the Riemann problem*

Given the initial states $\mathbf{u}_L$ and $\mathbf{u}_R$, the solution of the Riemann problem is complete if the intermediate state $\mathbf{u}_M$ is known. Recalling that $\mathbf{u}_M$ is connected to $\mathbf{u}_L$ by a wave (shock or rarefaction) contained into the first characteristic field, and to $\mathbf{u}_R$ by a wave (shock or rarefaction) contained into the second characteristic field, the intermediate state can be easily found at the intersection of the curves $H_1(\mathbf{u}, \mathbf{u}_L)$ and $H_2^B(\mathbf{u}, \mathbf{u}_R)$. This is equivalent to solving a nonlinear system of two scalar equations, where the unknowns are the components of $\mathbf{u}_M$. It is possible to demonstrate the following

**Proposition 5**. *The solution of the Riemann problem for the LInA model on horizontal bed exists if and only if the condition*

(21) $C: h_L u_L + 2h_L\sqrt{gh_L}/3 > h_R u_R - 2h_R\sqrt{gh_R}/3$

*is satisfied. If the solution exists, it is unique.*

*Proof.* The curve $H_1(\mathbf{u},\mathbf{u}_L)$ is strictly decreasing, and its supremum is

(22) $\lim_{h\to 0^+} hu = \lim_{h\to 0^+} h_L u_L + 2\left(h_L\sqrt{gh_L} - h\sqrt{gh}\right)/3 = h_L u_L + 2h_L\sqrt{gh_L}/3$.

Similarly, the curve $H_2^B(\mathbf{u},\mathbf{u}_R)$ is strictly increasing, and its infimum is

(23) $\lim_{h\to 0^+} hu = \lim_{h\to 0^+} h_R u_R - 2\left(h_R\sqrt{gh_R} - h\sqrt{gh}\right)/3 = h_R u_R - 2h_R\sqrt{gh_R}/3$.

This implies that an intersection of the curves $H_1(\mathbf{u},\mathbf{u}_L)$ and $H_2^B(\mathbf{u},\mathbf{u}_R)$ exists if and only if the condition of Eq. (21) is satisfied. From the Intermediate value Theorem, it follows that this intersection is unique.

An important consequence of *Proposition* 5 is that there is a class of initial conditions, namely those that do not satisfy the Eq. (21), for which the Riemann problem of the LInA Equations has no solution. This is a remarkable difference with respect to the SWE, where the Riemann problem admits a solution for any system of initial conditions (Toro 2001). An example with no solution for the LInA Riemann problem is represented in Figure 2a, with initial data in the first line of Table 1. The inspection of the panel shows that the curves $H_1(\mathbf{u},\mathbf{u}_L)$ and $H_2^B(\mathbf{u},\mathbf{u}_R)$ have no intersection, and this implies the absence of solution for the Riemann problem.

It is immediate to see that, depending on the relative position of $\mathbf{u}_L$ and $\mathbf{u}_R$ in the ($h$, $hu$) plane, the solution of the Riemann problem for the LInA equations may involve two rarefactions, two shocks, or a shock and a rarefaction. In Figure 2, the curves $H_1(\mathbf{u},\mathbf{u}_L)$ and $H_2^B(\mathbf{u},\mathbf{u}_R)$ are plotted for the initial conditions of Table 1, showing that different solution configurations are possible. In particular, an example with two rarefactions is represented in Figure 2b, while a shock-rarefaction and a two shock configuration are represented in Figure 2c and 2d, respectively. In all these cases, the numerical values of $h_M$ and $u_M$ can be found using the Newton-Raphson algorithm, whose convergence is ensured by the regularity of the curves $H_1(\mathbf{u},\mathbf{u}_L)$ and $H_2^B(\mathbf{u},\mathbf{u}_R)$.

It is important to note that the dam-break solutions presented in Martins et al. (2016a,b) are a sub-class (with $u_L = u_R = 0$ m/s) of the general Riemann problem solution presented here.

[Insert Figure 2 about here]

[insert Table 1 about here]

### *3.2 Riemann problem on a horizontal dry bed*

A comparison between the LInA equations and the SWE can be made by considering the solution of the Riemann problem on a dry bed, with $\mathbf{u}_R = \mathbf{0}$. Recalling the *Proposition* 5, it is evident that the LInA Riemann problem on the dry bed has no solution for $F(\mathbf{u}_L) \leq -2/3$, while the corresponding solution exists in the case of the SWE. For $F(\mathbf{u}_L) > -2/3$, the solution exists and the curve $H_2^B(\mathbf{u},\mathbf{u}_R)$ reduces to the curve $S_2^B(\mathbf{u},\mathbf{0})$ of Eq. (20).

It is possible to consider two classes of states $\mathbf{u}_L$ moving on the dry bed, namely the states with $F(\mathbf{u}_L) \in \left]-2/3, F_w\right]$ and the states with $F(\mathbf{u}_L) > F_w$. The states $\mathbf{u}_L$ characterized by

$F(\mathbf{u}_L) \in \,]-2/3, F_w]$ are represented in the ($h$, $hu$) plane by a point that lies below the curve $S_2^B(\mathbf{u},\mathbf{0})$. In this case, the intermediate state $\mathbf{u}_M$ between $\mathbf{u}_L$ and $\mathbf{u}_R = \mathbf{0}$ is connected to $\mathbf{u}_L$ by means of a rarefaction contained into the first characteristic field. An example solution (flow depth $h$) is represented with a continuous line in Figure 3a at time $t = 5$ s for the case $\mathbf{u}_L = (1.00 \quad 0.20)^T$. From the inspection of the figure, it is evident that the flow accelerates along the rarefaction curve from the state $\mathbf{u}_L$ to the state $\mathbf{u}_M$, which is characterized by $F(\mathbf{u}_M) = F_w$, while the state $\mathbf{u}_M$ is connected to the dry bed by a shock moving with speed $\sigma_2(\mathbf{u}_M,\mathbf{0}) = \sqrt{0.5gh_M}$. The slowest signal of this solution configuration is the left edge of the rarefaction wave, which moves with speed $\lambda_1(\mathbf{u}_L) = -\sqrt{gh_L}$. The solution of the SWE for the same problem is characterized by a rarefaction connecting the state $\mathbf{u}_L$ to the dry bed, and it is represented in Figure 3a with a dashed line. In the SWE solution, the slowest edge of the rarefaction wave moves with speed $\lambda_{SLOW} = u_L - \sqrt{gh_L}$, while the fastest edge of the rarefaction wave moves with speed $\lambda_{FAST} = u_L + 2\sqrt{gh_L}$.

When the state $\mathbf{u}_L$ is characterized by $F(\mathbf{u}_L) > F_w$, it lies above the curve $S_2^B(\mathbf{u},\mathbf{0})$ of the ($h$, $hu$) plane. In this case, the intermediate state $\mathbf{u}_M$ between $\mathbf{u}_L$ and $\mathbf{u}_R = \mathbf{0}$ is connected to $\mathbf{u}_L$ by a shock contained into the first characteristic field. An example solution (flow depth $h$) is represented in Figure 2b at time $t = 5$ s for the case $\mathbf{u}_L = (1.00 \quad 5.00)^T$. The inspection of the panel shows that the shock connecting the states $\mathbf{u}_L$ and $\mathbf{u}_M$ moves upstream with speed $\sigma_1(\mathbf{u}_M,\mathbf{u}_L) = -\sqrt{0.5g(h_M + h_L)}$, slowing down the state $\mathbf{u}_L$ to subcritical flow conditions characterized by $F(\mathbf{u}_M) = F_w$, while the downstream shock moves with speed $\sigma_2(\mathbf{u}_M,\mathbf{0}) = \sqrt{0.5gh_M}$. The appearance of the upstream shock is an unphysical feature of the solution that is not exhibited by the SWE, whose exact solution is plotted with a dashed line in the same panel.

It is useful to compare the slowest and the fastest wave speeds exhibited by LInA and SWE for the Riemann problem on dry bed. The coefficient $\rho_{SLOW}$ is the ratio between the slowest waves in the SWE and in the LInA equations solutions, and it is defined as

$$(24) \quad \rho_{SLOW} = \begin{cases} F(\mathbf{u}_L) \in \,]-2/3, F_w], & \left(u_L - \sqrt{gh_L}\right)/\lambda_1(\mathbf{u}_L) \\ F(\mathbf{u}_L) > F_w, & \left(u_L - \sqrt{gh_L}\right)/\sigma_1(\mathbf{u}_M, \mathbf{u}_L) \end{cases},$$

while $\rho_{FAST}$ is the ratio between the fastest waves in the SWE and in the LInA equations solutions, and it is defined as

$$(25) \quad \rho_{FAST} = \left(u_L + 2\sqrt{gh_L}\right)/\sigma_2(\mathbf{u}_M, \mathbf{0}).$$

The coefficients $\rho_{SLOW}$ and $\rho_{FAST}$ are plotted in Figure 4 for $F(\mathbf{u}_L) \in [-0.5, 2]$. The slowest wave in the LInA solution is similar to that of the SWE solution when $F(\mathbf{u}_L)$ is close to zero. In particular, the percentage error is less than 20% for $F(\mathbf{u}_L) \in [-0.2, 0.2]$, but it increases rapidly for $|F(\mathbf{u}_L)| > 0.2$. Very interestingly, both the slowest waves keep the negative sign for $F(\mathbf{u}_L) < 1$, while the sign of the slowest wave in the SWE solution becomes positive for supercritical flows. Finally, the figure shows that in the interval considered the fastest wave of the SWE solution is at least 3.83 times greater than the fastest wave of the LInA solution. In other words, the LInA equations strongly underestimate the speed of the wetting phenomenon on a frictionless horizontal dry bed.

[Insert Figure 3 about here]

[Insert Figure 4 about here]

## 3.3 Impact on a wall

The impact on a wall can be simulated by considering a Riemann problem where $h_R = h_L$, and $u_R = -u_L$ (Toro 2001). In Figure 5a, the exact solution (flow depth) of the LInA equations is represented with a continuous line at time $t = 5$ s for the case $\mathbf{u}_L = (1.00 \quad 1.50)^T$. The inspection of the panel shows that the impact on the wall causes the increase of the flow depth, and the formation of a shock moving upstream that decelerates the flow to the rest. In the same panel, the solution for the SWE is represented with a dashed line. The comparison between the plots shows that the SWE produce a higher flow depth at the wall, while the speed of the backward shock is smaller.

The ratio $h_{LIna}/h_{SWE}$ between the wall flow depths supplied by the two mathematical models is represented in Figure 5b for different values of the incoming flow Froude number $F(\mathbf{u}_L)$. From the figure, it is evident that the error is greater than 5% for $F(\mathbf{u}_L) > 0.45$, and it is greater than 10% for $F(\mathbf{u}_L) > 0.70$.

[Insert Figure 5 about here]

## 4 Exact solution for floodplain wetting with friction

In Section 2 it has been shown that the solutions of the LInA model with a wet-dry frontier always exhibits a shock, also when starting from smooth initial conditions, and this shock has been characterized in Section 3. In the present section it is demonstrated how the LInA solution corresponding to a nonbreaking wave on a horizontal bed with uniform velocity can be constructed.

Since $S_0 = 0$, the analysis of Subsection 2.1 is repeated in dimensional form, and the LInA system of Eq. (2) is solved with initial flow depth $h_0(x) = h(x,0)$ and constant uniform flow velocity

$u(x,t) = u_0 > 0$. If the Chézy formula is used for the calculation of the friction slope $S_f$, the LInA model can be simplified as follows:

$$(26) \quad \begin{aligned} \frac{\partial h}{\partial t} + u_0 \frac{\partial h}{\partial x} &= 0 \\ u_0 \frac{\partial h}{\partial t} + gh \frac{\partial h}{\partial x} &= -g \frac{u_0^2}{C^2} \end{aligned},$$

where $C$ is the Chézy coefficient. It is evident that the first of Eq. (26) admits the exact solution

$$(27) \quad h(x,t) = h_0(x - u_0 t),$$

which implies that the flow profile translates rigidly with velocity $u_0$. If $h_0'(x)$ is the derivative of the initial flow depth $h_0(x)$, the second of Eq. (26) can be rewritten as an ordinary differential equation, as follows:

$$(28) \quad (gh_0 - u_0^2) h_0' = -g \frac{u_0^2}{C^2}.$$

From Eq. (28), it is confirmed that the flow profile cannot be continuously connected to the dry bed. In fact, the flow profile slope $h_0'$ corresponding to $h_0 = 0$ is positive, and this configuration must be excluded because it implies that the flow-depth solution is multivalued at the wet-dry frontier (see Sub-section 2.1). The only remaining possibility is that the smooth part of the flow profile, with $h_0' < 0$, is connected to the dry bed by a shock that is contained into the second characteristic field and whose celerity is $\sigma_2 = u_0$. From Eq. (18), it follows that the flow behind this shock is characterized by flow depth $h_F = (2 u_0^2)/g$ and Froude number $F_F = u_0 / \sqrt{gh_F} = F_w$. If the initial abscissa corresponding to the shock is $x_F$, the initial flow profile slope corresponding to the depth $h_F = h_0(x_F)$ is $\lim_{x \to x_F} h_0'(x) = -g/C^2$, and this confirms that the initial flow depth $h_0$ increases from downstream to upstream with minimum flow depth $h_F$ at the wetting front.

The Eq. (28) can be solved by separation of variables, obtaining

$$(29) \quad h_0(x) = \begin{cases} x < x_F, & h_B\left[F_B^2 + \sqrt{(1-F_B^2)^2 - \dfrac{x}{x_0}}\right], \\ x > x_F, & 0 \end{cases}$$

where $x_0 = C^2 h_B / (2gF_B^2)$, $h_B = h_0(0)$ is the flow depth at the reference abscissa $x = 0$ and $F_B = u_0 / \sqrt{gh_B}$ is the corresponding Froude number, while the shock position $x_F$ is defined by

$$(30) \quad x_F = x_0 - \frac{C^2 h_B}{g}.$$

The bound $h_B \geq h_F$ implies that only the subcritical reference conditions characterized by $F_B \leq F_w$ can be imposed to the LInA model for this problem. The application of a similar procedure to the SWE of Eq. (1) leads to the solution

$$(31) \quad h_0(x) = \begin{cases} x \leq x_0, & h_B\sqrt{1 - \dfrac{x}{x_0}} \\ x > x_0, & 0 \end{cases}.$$

No additional bound is imposed to $h_B > 0$ in Eq. (31), and this means that the flow profile corresponding to the SWE can be constructed not only for $F_B \leq F_w$, but also for the case $F_B > F_w$. As expected from *Proposition* 1, the shock at the front of the LInA equations solution disappears in the limit $F_B \to 0$, where Eqs. (29) and (31) coincide.

The flow profile of Eq. (29) can be represented in the plane $\left(\xi, h_0/h_B\right)$, where the dimensionless distance is defined by $\xi = x/x_0$. The cases $F_B = 0.3$ (thick dashed line) and $F_B = 0.5$ (thick solid line) are plotted in the Figure 6, where the flow profile of Eq. (31) corresponding to the SWE is also represented with a thin solid line. The inspection of Figure 6 shows that the LInA model exhibits a well developed shock at the front, and the solution corresponding to $F_B = 0.3$ better approximates the SWE solution than the case with $F_B = 0.5$, which is closer to the limit $F_w$.

[Insert Figure 6 about here]

**5 Solution of the Riemann problem on uneven bed**

Geometric singularities such as terrain steps and drops, sills, embankment and levee sides, channel sidewalls, and sidewalk steps, are commonly present in the flooded areas. These features, which are all characterized by a rapid variation of the terrain elevation, can be modeled by bed elevation discontinuities (Aronica et al. 1998, Cozzolino et al. 2011). Recalling that the leading edge of flooding waves in the LInA model is always represented by a discontinuity moving on dry bed (see Sections 2 and 4), it is interesting to consider the interaction between such a discontinuity and the bed elevation discontinuities. This task can be properly tackled in the context of an augmented Riemann problem, where the bed elevation is added among the conserved variables.

The systematic analysis of the Riemann problem on uneven bed in the LInA model is beyond the scope of the present paper, and for this reason only some example will be considered. The corresponding results will be compared with those supplied by the SWE, for which the solution of the discontinuous-bed Riemann problem is discussed in Alcrudo and Benkhaldoun (2001).

*5.1 Position of the Riemann problem and general solution*

It is possible to consider an augmented system of partial differential equations obtained by adding the trivial equation $\partial z/\partial t = 0$ to the system of Eq. (1). If the friction is absent, the augmented system is rewritten as

$$(32) \quad \frac{\partial \mathbf{U}}{\partial t} + \frac{\partial \mathbf{F}(\mathbf{U})}{\partial x} + \mathbf{H}(\mathbf{U})\frac{\partial \mathbf{U}}{\partial x} = 0,$$

where $\mathbf{U}(x,t) = \begin{pmatrix} h & hu & z \end{pmatrix}^T$ is the augmented vector of the conserved variables, $\mathbf{F}(\mathbf{U}) = \begin{pmatrix} hu & gh^2/2 & 0 \end{pmatrix}^T$ is the augmented vector of the fluxes, while the matrix $\mathbf{H}(\mathbf{U})$ is defined as

$$(33) \quad \mathbf{H}(\mathbf{U}) = \begin{pmatrix} 0 & 0 & 0 \\ 0 & 0 & gh \\ 0 & 0 & 0 \end{pmatrix}.$$

It is easy to see that the matrix $\mathbf{A}(\mathbf{U}) = \partial \mathbf{F}/\partial \mathbf{U} + \mathbf{H}(\mathbf{U})$ has the following three distinct real eigenvalues

$$(34) \quad \lambda_1(\mathbf{U}) = -\sqrt{gh}, \quad \lambda_0(\mathbf{U}) = 0, \quad \lambda_2(\mathbf{U}) = \sqrt{gh},$$

with corresponding linearly independent eigenvectors

$$(35) \quad \mathbf{r}_1(\mathbf{U}) = \begin{pmatrix} 1 & -\sqrt{gh} & 0 \end{pmatrix}^T, \quad \mathbf{r}_0(\mathbf{U}) = \begin{pmatrix} 1 & 0 & -1 \end{pmatrix}^T, \quad \mathbf{r}_2(\mathbf{U}) = \begin{pmatrix} 1 & \sqrt{gh} & 0 \end{pmatrix}^T.$$

It is evident that the system of Eq. (32) is strictly hyperbolic, and that it differs from the system of Eq. (13) because the conserved variable $z$ has introduced an additional characteristic field. The same arguments used in Appendix A for the system of Eq. (13) can be used to show that the characteristic fields related to the eigenvectors $\mathbf{r}_1(\mathbf{U})$ and $\mathbf{r}_2(\mathbf{U})$ are genuinely non-linear, and that they may contain moving waves such as shocks or rarefactions. Interestingly, the third component of the vectors $\mathbf{r}_1(\mathbf{U})$ and $\mathbf{r}_2(\mathbf{U})$ is null, and this means that the shocks and the rarefactions contained in the corresponding characteristic fields may develop only where the bed elevation is constant. The comparison between Eq. (35) and Eq. (14) leads to the conclusion that the projection on the ($h$, $hu$) plane of 1- and 2-waves coincides with the expressions of Eqs. (16) and (17).

Special attention should be payed to the characteristic field corresponding to the eigenvector $\mathbf{r}_0(\mathbf{U})$. It is immediate to see that this characteristic field is linearly degenerate, because $\nabla^T \lambda_0(\mathbf{U}) \mathbf{r}_0(\mathbf{U}) = 0$, and this means that the 0-waves contained into this characteristic field are special discontinuities called contact discontinuities.

The Riemann problem consists in solving Eq. (32) with initial conditions

(36) $\mathbf{U}(x,0) = \begin{cases} x<0, & \mathbf{U}_L \\ x>0, & \mathbf{U}_R \end{cases}$,

where $\mathbf{U}_L = \begin{pmatrix} h_L & h_L u_L & z_L \end{pmatrix}^T$ and $\mathbf{U}_R = \begin{pmatrix} h_R & h_R u_R & z_R \end{pmatrix}^T$. Also in this case, the solution of the Riemann problem is self-similar, and a vector function $\mathbf{V}(s)$ of the scalar parameter $s$ exists such that $\mathbf{U}(x,t) = \mathbf{V}(x/t)$. Recalling that $\lambda_1(\mathbf{U}) < \lambda_0(\mathbf{U}) < \lambda_2(\mathbf{U})$, it is immediate to see that the solution of the Riemann problem consists in the four ordered states $\mathbf{U}_L$, $\mathbf{U}_1 = \begin{pmatrix} h_1 & h_1 u_1 & z_L \end{pmatrix}^T$, $\mathbf{U}_2 = \begin{pmatrix} h_2 & h_2 u_2 & z_R \end{pmatrix}^T$, and $\mathbf{U}_R$, where $\mathbf{U}_L$ and $\mathbf{U}_1$ are connected by a 1-wave, $\mathbf{U}_1$ and $\mathbf{U}_2$ are connected by a 0-wave, while $\mathbf{U}_2$ and $\mathbf{U}_R$ are connected by a 2-wave. In the following, the symbols

$\mathbf{u}_L = \begin{pmatrix} h_L & h_L u_L \end{pmatrix}^T$, $\mathbf{u}_R = \begin{pmatrix} h_R & h_R u_R \end{pmatrix}^T$, $\mathbf{u}_1 = \begin{pmatrix} h_1 & h_1 u_1 \end{pmatrix}^T$, and $\mathbf{u}_2 = \begin{pmatrix} h_2 & h_2 u_2 \end{pmatrix}^T$, are used to indicate the projections on the ($h$, $hu$) plane of the vectors $\mathbf{U}_L$, $\mathbf{U}_R$, $\mathbf{U}_1$, and $\mathbf{U}_2$, respectively.

*5.2 Definition of the contact discontinuity*

It is easy to see that the speed $\sigma_0(\mathbf{U}_1, \mathbf{U}_2)$ of the discontinuities contained into the characteristic field associated to $\mathbf{r}_0(\mathbf{U})$ is null. In fact, the application of the Rankine-Hugoniot condition (LeVeque 1992) to the trivial equation $\partial z/\partial t = 0$ leads to

(37) $(z_R - z_L)\sigma_0(\mathbf{U}_1, \mathbf{U}_2) = 0$,

because the corresponding flux is identically null. The Eq. (37) implies that the speed $\sigma_0(\mathbf{U}_1, \mathbf{U}_2)$ is null for $z_R - z_L \neq 0$, and that the constant states $\mathbf{U}_1$ and $\mathbf{U}_2$ connected by the 0-wave are located to the left and to the right of $x = 0$, respectively.

The mathematical relationship between $\mathbf{U}_1$ and $\mathbf{U}_2$ can be constructed by observing that the characteristic field corresponding to the eigenvector $\mathbf{r}_0(\mathbf{U})$ can be associated to the steady flow solutions of Eq. (32). In fact, these solutions satisfy the condition

(38) $\mathbf{A}(\mathbf{U})\dfrac{\partial \mathbf{U}}{\partial x} = 0$,

which means that $\partial \mathbf{U}/\partial x$ must be proportional to an eigenvector of $\mathbf{A}$ with identically null eigenvalue. This eigenvector is precisely $\mathbf{r}_0(\mathbf{U})$, and the proportionality equation

(39) $$\frac{\partial \mathbf{U}}{\partial x} = \alpha(x)\mathbf{r}_0(\mathbf{U}(x)),$$

is a differential equation with solution $z + h$ = const. and $hu$ = const. (see Appendix E). It is easy to verify that this solution remains meaningful in the case of bed elevation discontinuity, and this suggests to define the 0-wave as

(40) $$H_0(\mathbf{U}_2, \mathbf{U}_1) = \begin{cases} h_1 u_1 = h_2 u_2 \\ z_L + h_1 = z_R + h_2 \end{cases}.$$

The 0-wave definition of Eq. (40) coincides with the choice made by Aronica et al. (1998) for the inner boundary conditions characterized by rapid variations of the bed elevation.

*5.3 Riemann problem at the dry bed step*

The Riemann problem at the dry bed step consists in solving the system of Eq. (32) with initial conditions of Eq. (36), where $h_R = 0$ m, $h_R u_R = 0$ m²/s, and $z_L < z_R$. The problem can be solved by finding the projections $\mathbf{u}_1$ and $\mathbf{u}_2$ of $\mathbf{U}_1$ and $\mathbf{U}_2$ on the plane $(h, hu)$, respectively, that satisfy the Eq. (40) and that lie on the curves $H_1(\mathbf{u}, \mathbf{u}_L)$ and $H_2^B(\mathbf{u}, \mathbf{0})$, respectively. Instead of doing a systematic analysis of all the possible solution configurations, as made in Section 2 for the Riemann problem on horizontal bed, some example solution will be considered in the present subsection, in order to discuss the differences between the LInA model and the SWE on uneven bed. The initial data for the examples considered are resumed in Table 2, while the corresponding solutions at time $t = 5$ s (free surface elevation) are plotted in Figure 7.

The panel of Figure 7a (continuous line) corresponds to the case where the incoming flow is subcritical ($F(\mathbf{u}_L) = 0.064 < F_w$), and the corresponding energy is higher than that strictly required to pass over the obstacle. The state $\mathbf{U}_L$ is connected to the state $\mathbf{U}_1$ by a rarefaction, while the state $\mathbf{U}_2$ over the bed step is connected to the dry bed on the step by a shock. The Froude number corresponding to the flow moving on the dry bed is fixed, and equal to $F(\mathbf{u}_2) = F_w$, as shown in Subsection 3.1.2. The solution of the SWE for the same problem is represented with a dashed line, and exhibits a right rarefaction whose celerity is greater than the celerity of the fastest signal in the LInA equations. Notably, the underestimation of the fastest signal does not correspond to an underestimation of the discharge over the bed step, because the discharge per unit width through the geometric discontinuity exhibited by the LInA equations is $h_1 u_1 = 0.537$ m²/s, and it is higher than the value $h_1 u_1 = 0.459$ m²/s corresponding to the Shallow water Equations, while the flood stage immediately to the left of the obstacle is similar. If $H(\mathbf{U}) = h + u^2/(2g) + z$ is the total head related to the generic state $\mathbf{U}$, it is immediate to see that $H(\mathbf{U}_1) - H(\mathbf{U}_2) = -0.079$ m, meaning that the energy of the flow is increased through the geometric discontinuity. In absence of external devices such as pumps, this phenomenon is clearly unphysical.

The Figure 7b represents a subcritical case (initial data in Table 2, with $F(\mathbf{u}_L) = 0.479 < F_w$) where the bed step in the LInA Riemann problem is high enough to completely reflect the incoming flow from the left, leaving the top of the bed step dry. The SWE exhibit a totally different behavior (dashed line in Figure 7b) for the same data, because the energy is sufficient to make the flow jump over the bed step.

In Figure 7c it is represented the case where the incoming flow is supercritical (initial data in Table 2, with $F(\mathbf{u}_L) = 1.596 > F_w$) and the dry bed step is high enough to partially reflect the flow ($z_R - z_L = 0.50$ m). For this reason, the solution of the LInA equations exhibit a fast shock that moves backward, while a slower shock propagates over the dry bed step. The flow depth immediately to the left of the geometric discontinuity is $h_1 = 1.640$ m, while the discharge passing through the geometric discontinuity is $h_1 u_1 = 2.696$ m²/s. On the contrary, the Shallow water Equations (dashed line) exhibit a shock that propagates backward very slowly, but a fast rarefaction propagates on the dry step. In the example of Figure 7c, the LInA model underestimates not only the flooded area, but also the discharge through the geometric discontinuity and the flood stage, because the SWE predict $h_1u_1$ = 4.203 m²/s and $h_1$ = 2.126 m. In addition, the LInA model predicts an increase of energy through the geometric discontinuity, because $H(\mathbf{U}_1) - H(\mathbf{U}_2)$ = - 0.147 m.

The example solution of Figure 7d can be instructively compared with that of Figure 7c, because the characteristics of the incoming flow are identical, while the bed step is significantly lower ($z_R - z_L$ = 0.10 m). In this case, the LInA equations predict again the partial reflection of the flow against of the obstacle (continuous line), with flow depth $h_1$ = 1.446 m and discharge $h_1u_1$ = 3.457 m²/s. The structure of the solution corresponding to the Shallow water Equations (dashed line) is very different, because it corresponds to the complete transmission of the incoming wave, which has sufficient energy to jump over the bed step without reflection. In particular, $h_1 = h_L$, while the discharge over the bed step predicted by the Shallow water Equations is $h_1u_1$ = 5 m²/s. This confirms that the Shallow water Equations are able to make the flow jump over obstacles that reflect partially or totally the waves computed by the LInA equations. Also in this case, the LInA equations predict an increase of energy through the geometric discontinuity, because $H(\mathbf{U}_1) - H(\mathbf{U}_2)$ = - 0.045 m.

[Insert Figure 7 about here]

[insert Table 2 about here]

*5.4 Riemann problem at the dry bed drop*

When the Riemann problem at the bed drop is considered, the system of Eq. (32) is solved with the initial conditions of Eq. (36), where $h_R$ = 0 m, $h_Ru_R$ = 0 m²/s, and $z_L > z_R$. The initial data for the examples considered in the present subsection are reported in Table 3, while the corresponding solution at time $t$ = 5 s (free surface elevation) is plotted in Figure 8.

In Figure 8a, the solution of the Riemann problem is represented for the same flow conditions of Figure 7a, (subcritical incoming flow with $F(\mathbf{u}_L) = 0.064 < F_w$), but now there is a bed drop ($z_R - z_L$ = - 0.20 m) instead of a bed step. The incoming flow $\mathbf{U}_L$ is connected to the state $\mathbf{U}_1$ to the left of the geometric discontinuity by a rarefaction, while the state $\mathbf{U}_2$ to the right of the geometric discontinuity is connected to the dry bed on the step by a shock. The flow $\mathbf{U}_1$ at the top of the bed drop is expected to be critical, but the LInA equations supply a supercritical flow, characterized by Froude number $F(\mathbf{u}_1) = 1.126$ and velocity $u_1 = 2.615$ m/s, while the Froude number corresponding to the flow moving on the dry bed is fixed, and equal to $F(\mathbf{u}_2) = F_w$. The solution of the SWEs for the same problem is represented with a dashed line, and consists of a rarefaction that accelerates the incoming flow until the critical conditions are attained immediately to the left of the bed drop ($F(\mathbf{u}_1) = 1$), while the supercritical flow ($F(\mathbf{u}_2) = 2.052$) at the foot of the bed drop is connected to the dry bed by a rarefaction. The LinA equations underestimate the fastest signal, but overestimate the discharge passing through the geometric discontinuity ($h_1 u_1 = 1.437$ m²/s instead of $h_1 u_1 = 1.012$ m²/s). Interestingly, $H(\mathbf{U}_1) - H(\mathbf{U}_2) = 0.161$ m, meaning that the energy of the flow is loss through the geometric discontinuity.

In the case of Figure 8b, the characteristics of the incoming flow coincide with those of Figure 8a, but now the bed drop is higher ($z_R - z_L = -1.01$ m). Despite the similarity of the solution structure with that of Figure 8a, the velocity associated to the state $\mathbf{U}_1$ is now $u_1 = 203.760$ m/s, with flow depth $h_1 = 0.011$ m, and this means that the velocity of supercritical flows at the top of the bed drop depends on the height of the drop, which is clearly unphysical. This aspect will be examined in the next subsection. Due to the high value of the flow velocity $u_1$, the head $H(\mathbf{U}_1) = 2116.134$ m is out of the range of values encountered in practical applications, and the head loss $H(\mathbf{U}_1) - H(\mathbf{U}_2) = 2115.867$ m at the geometric discontinuity is correspondingly high.

In Figure 8c, the subcritical incoming flow ($F(\mathbf{u}_L) = 0.836 > F_w$) encounters a small bed drop. The classic SWE supply a solution (dashed line) where the flow is accelerated towards the critical state at the top of the bed drop by means of a rarefaction, while the flow is supercritical at the step foot. On the contrary, the LInA model supplies a solution configuration where the flow is decelerated by a shock connecting the states $\mathbf{U}_L$ and $\mathbf{U}_1$ that moves backward, while the flow $\mathbf{U}_2$ at the step foot is subcritical. The head loss $H(\mathbf{U}_1) - H(\mathbf{U}_2) = 0.005$ m at the geometric discontinuity is very low, in this case.

A case where the incoming flow is supercritical ($F(\mathbf{u}_L) = 1.596 > F_w$) is represented in Figure 8d. The SWE supply a solution where this flow is not influenced by the presence of the bed drop, while the solution corresponding to the LInA model exhibits a rarefaction that propagates upstream. The head loss through the geometric discontinuity is $H(\mathbf{U}_1) - H(\mathbf{U}_2) = 1.759$ m, and it is quite high.

[Insert Figure 8 about here]

[insert Table 3 about here]

*5.5 Limits of existence for the Riemann problem at the dry bed drop*

The case of Figure 8b seems very close to the applicability limits of the LInA model because the water depth at the top of the bed drop is close to zero, while the velocity is very high and out of the range of practical applications. These observations stimulate a closer analysis of the Riemann problem at the dry bed drop, in order to plot a diagram with the fields of existence of the different types of solution.

Preliminarily, it is observed that the Froude number of the flow at the top of the bed drop is constrained by the drop height. In fact, it is immediate to see that the projection $\mathbf{u}_2$ of $\mathbf{U}_2$ on the ($h$, $hu$) plane must lie on the curve $S_2^B(\mathbf{u}, \mathbf{0})$, while the states $\mathbf{U}_1$ and $\mathbf{U}_2$ must satisfy the Eq. (40). The two conditions lead to the solution

(41) $F(\mathbf{u}_1) = F_w \left(1 + \dfrac{a}{h_1}\right)^{3/2}$,

where $a = z_L - z_R > 0$ is the step height. It is immediate to see that that a shock connects the state $\mathbf{U}_L$ to the state $\mathbf{U}_1$ when $F(\mathbf{u}_L) > F(\mathbf{u}_1)$, while a rarefaction connects the state $\mathbf{U}_L$ to the state $\mathbf{U}_1$ when $F(\mathbf{u}_L) < F(\mathbf{u}_1)$. The position $\mathbf{U}_L = \mathbf{U}_1$ leads to the limiting condition

$$(42) \quad F(\mathbf{u}_L) = F_w \left(1 + \frac{a}{h_L}\right)^{3/2}$$

between the solutions with a shock and the solutions with a rarefaction. The Eq. (42) is represented in the plane ($F(\mathbf{u}_L)$, $a/h_L$) of Figure 9 with a dashed line.

If the left Froude number $F(\mathbf{u}_L)$ is less than the limit of Eq. (42) for a given value of $a/h_L$, a rarefaction contained in the first characteristic field connects the states $\mathbf{U}_L$ and $\mathbf{U}_1$, and this implies that $\mathbf{u}_1$ lies on the curve curve $R_1(\mathbf{u}, \mathbf{u}_L)$. After the position $\mathbf{u} = \mathbf{u}_1$ and $\mathbf{u}_0 = \mathbf{u}_L$ in Eq. (16), and the substitution in Eq. (41), some algebra leads to

$$(43) \quad F_w \left(\frac{h_1}{h_L} + \frac{a}{h_L}\right)^{3/2} = F(\mathbf{u}_L) + \frac{2}{3}\left[1 - \left(\frac{h_1}{h_L}\right)^{3/2}\right]$$

which makes sense only for $h_1 > 0$. The limit $h_1 \to 0$ leads to the condition

$$(44) \quad F(\mathbf{u}_L) = F_w \left(\frac{a}{h_L}\right)^{3/2} - \frac{2}{3},$$

which is represented in Figure 9 with a continuous line. The initial conditions which are close to the limit of Eq. (44) lead to arbitrarily high values of the velocity $u_1$, and this explains the results of the Riemann problem depicted in Figure 8b. The region above the curve of Eq. (44) is forbidden, and this implies that there are initial conditions for the Riemann problem at the dry bed drop that do not admit any solution. This is in contrast with the SWE, where the solution of the Riemann problem at bed elevation discontinuities always exists.

[Insert Figure 9 about here]

# 6 Numerical experiments

In this section, the characteristics of a widely used numerical model for the approximate solution of the LInA, namely the *q-centred* finite-difference scheme by de Almeida et al. (2012), are scrutinized with reference to the flood propagation computation. Due to its incomplete conservativity, this scheme is not able to capture the frontal shock exhibited by the LInA waves that advance on dry bed. For this reason, a fully conservative Finite Volume scheme is additionally considered. The choice of the Finite Volume method is dictated by the observation that this type of numerical schemes is now commonly used in research and commercial flooding models (Toro 2001), due to the conservativity properties, the ability to capture traveling shocks and wet-dry frontiers, the grid flexibility, and the availability of theorems for the convergence of the solutions (Eymard et al. 2007). The Rusanov Riemann solver (Toro 2009) is chosen here for the calculation of the numerical fluxes because it ensures the depth-positivity by interface under an appropriate time restriction (Bouchut 2004) when it is applied to the SWE, and it is less computationally expensive than the HLL approximate Riemann solver.

Due to the lack of mass conservation, both the *q-centred* and the Finite Volume scheme must be modified in order to enforce the depth-positivity by dynamically reducing the time step.

## *6.1 The q-centred scheme by de Almeida et al. (2012)*

In the *q-centred* scheme (de Almeida et al. 2012), which is an improvement of the staggered-grid finite-difference scheme by Bates et al. (2010), the flow field is subdivided into rectangular cells of length $\Delta x$. The variables $z_i$, $h_i$, and $\zeta_i = h_i + z_i$ (free surface elevation), are stored at the center of the $i$-th cell, while the discharge $q_{i-1/2}$ is stored at the interface between the cells $i$-1 and $i$. After having rewritten the second of Eq. (2) in the non-conservative form

(45) $\dfrac{\partial q}{\partial t} + gh\dfrac{\partial h}{\partial x} = -gh\dfrac{\partial z}{\partial x} - ghS_f$,

where $q = hu$, the discharge $q_{i-1/2}$ is advanced from the time level $t^n$ to the time level $t^{n+1}$ by means of the following non-conservative scheme:

(46) $q_{i-1/2}^{n+1} = \theta q_{i-1/2}^n + \dfrac{1-\theta}{2}\left(q_{i-3/2}^n + q_{i+1/2}^n\right) - \lambda g h_{f,i-1/2}^n \left(\zeta_i^n - \zeta_{i-1}^n\right)$.

where $h_{f,i-1/2}^n = \max\left[\zeta_{i-1}^n, \zeta_i^n\right] - \max\left[z_{i-1}, z_i\right]$, $\theta \in [0,1)$ is an user-defined coefficient, $\Delta t = t^{n+1} - t^n$ is the time step, and $\lambda = \Delta t/\Delta x$. Once that the discharge at the interfaces has been adjourned, the free-surface elevation in the cells is advanced in time by means of the conservative scheme

(47) $h_i^{n+1} = h_i^n - \lambda\left(q_{i+1/2}^{n+1} - q_{i-1/2}^{n+1}\right)$.

If $h_i^{n+1} < 0$, the position $h_i^{n+1} = 0$ is made: in this case, the global mass conservation is violated but $h_i^{n+1} \geq 0$ everywhere. For linear stability requirements, the time step $\Delta t$ used at the time level $t^n$ satisfies the inequality $\Delta t \leq \Delta t_0$, where

$$\text{(48)} \quad \Delta t_0 = \min\left[\Delta t_{max}, \alpha \frac{\Delta x}{\sqrt{gh^n_{max}}}\right].$$

In Eq. (48), $h^n_{max} = \max_i\{h^n_i\}$ is the maximum flow depth within the computational domain at the time level $t^n$, $\alpha \in (0,1]$ is an user defined coefficient, and $\Delta t_{max}$ is the maximum allowed time step for accuracy purposes. In the following, the coefficients $\alpha = 0.7$ (de Almeida et al. 2012), $\theta = 0.7$ (de Almeida and Bates 2013), $\Delta t_{max} = 0.1$ s, and $\Delta t = \Delta t_0$ are constantly used.

*6.1.1 Riemann problems on flat bed (Tests 1 and 2)*

The *q-centred* scheme by de Almeida et al. (2012) is applied to two Riemann problems on flat dry bed. The first Riemann problem, called Test 1, is characterized by $\mathbf{u}_L = (1.00 \quad 5.00)^T$ and $\mathbf{u}_R = (0 \quad 0)^T$, and the results of the calculations corresponding to the channel with length $L = 100$ m ($\Delta x = 0.02$) are represented for $t = 5.002$ s in Figure 10a with a dashed line (flow depth). The comparison with the exact solution (continuous thick line) shows that the celerity and the strength of the shocks are only approximately captured, confirming that numerical schemes written in non-conservative form cannot converge to the exact solution when discontinuities are present (Hou and LeFloch 1994). In particular, the *q-centred* scheme is not able to capture the strength and the propagation celerity of the wetting front, and the error about the corresponding position increases with time. During the simulation, the theoretical variation of water volume contained in the channel can be calculated with $\Delta W(t) = h_L u_L t$, and this supplies the theoretical volume of water contained into the physical domain at the end of calculations. In the present case, the relative difference between the computed final volume and the theoretical final volume is $5.68 \cdot 10^{-16}$, which is close to the machine error.

The second Riemann problem, called Test 2, is characterized by $\mathbf{u}_L = \begin{pmatrix} 1.00 & -5.00 \end{pmatrix}^T$ and $\mathbf{u}_R = \begin{pmatrix} 0 & 0 \end{pmatrix}^T$. Despite the fact that the LInA equations do not admit an exact solution for this Riemann problem on the dry bed ($F(\mathbf{u}_L) \leq -2/3$), the numerical model computes a solution whose results are represented with a dashed line in Figure 10b for $t$ = 5.002 s. The inspection of the figure shows that this solution consists of a receding flow which moves in the form of a rarefaction wave. It is evident that this feature, which is apparently of physical nature, is a purely numerical artifact. In the channel of length $L$ = 100 m, the relative difference between the computed final volume and the theoretical final volume is 0.578, which is far out of the admissible range for the practical applications. The comparison with the preceding Riemann problem suggests that errors in mass conservation are connected to the presence of receding flows, which cannot be simulated by the LInA mathematical model.

[Insert Figure 10 about here]

*6.1.2 Oscillations in the parabolic channel (Test 3)*

The Test 3 consists of the oscillations of the SWE model in a one-dimensional frictionless rectangular channel with bed elevation described by

(49) $\quad z(x) = z_0 \left( \dfrac{x}{a} \right)^2,$

where $a$, and $z_0$ are parameters. This case is particularly interesting because it combines advancing fronts, receding fronts, and a non-trivial bed elevation. At time $t$ = 0 s, the initial velocity is null, while the initial flow depth is described by

(50) $h(x,0) = \max\left[0, \zeta_0 - \dfrac{u_{max}\omega}{g}x - z(x)\right],$

where $u_{max}$ and $\zeta_0$ are parameters, while $\omega = \sqrt{2z_0 g}/a$. For $t > 0$ s, the mass oscillates with period $T = 2\pi/\omega$, maintaining a planar free surface profile. The exact solution for the flow depth is described by (Thacker 1981)

(51) $h(x,t) = \max\left[0, \zeta_0 - \dfrac{u_{max}\omega}{g}x\cos\omega t + \dfrac{u_{max}^2}{2g}\left(1 - \cos^2\omega t\right) - z(x)\right],$

and the maximum flow depth through the domain is $h_{max} = \zeta_0 + u_{max}^2/(2g)$.

The *q-centred* scheme by de Almeida et al. (2012) is applied with $\Delta x = 0.12$ m to the case with parameters $a$ = 2000 m, $z_0$ = 4 m, $\zeta_0$ = 3 m, $u_{max}$ = 2 m/s, and the corresponding results are represented in Figure 11 for the times $t_1$ = 350.25 s (Figure 11a), $t_2$ = 710.52 s (Figure 11b), $t_3$ = 1060.79 s (Figure 11c), and $t_4$ = 1421.02 s (Figure 11d). In the same figure, the exact solution of the SWE is also represented. The inspection of the diagram shows that there is no general agreement between the LInA *q-centred* numerical solution and the SWE exact solution, because the average free surface slope is not well captured, together with the position of the wetting-drying fronts. In contrast with the SWE exact solution, the Figures 11a and 11b clearly show the LInA shock at the advancing head (right front) of the oscillating water body. Interestingly, the *q-centred* scheme by de Almeida et al. (2012) supplies receding wet-dry frontiers (left front) that are continuously connected to the dry bed, and this feature is purely numerical, because the LInA model does not admit a receding front.

The maximum value of the flow depth supplied by the numerical model during the calculations is $h_{max}$ = 4.15 m, and it is 29% greater than the exact value. The main reason of this discrepancy seems the lack of mass conservation by the numerical algorithm. For example, the relative difference between the numerical and exact volume at time $t_4$ is 0.428, and this is clearly unacceptable. Corresponding to the value calculated for $h_{max}$, the minimum time step used by the algorithm is $\Delta t_{min} = 0.0132$ s.

[Insert Figure 11 about here]

### *6.2 A modified q-centred scheme*

The preceding calculations prompt a modification of the scheme by de Almeida et al. (2012) in order to preserve the mass conservation at every step of the algorithm. Aiming at this, we observe that the substitution of Eq. (46) into Eq. (47) leads to

$$
(52) \quad h_i^{n+1} = h_i^n - \frac{\lambda}{2}\left[(3\theta-1)\left(q_{i+1/2}^n - q_{i-1/2}^n\right) + (1-\theta)\left(q_{i+3/2}^n - q_{i-3/2}^n\right)\right] + \\
-\lambda^2\left[gh_{f,i+1/2}^n\left(\zeta_{i+1}^n - \zeta_i^n\right) - gh_{f,i-1/2}^n\left(\zeta_i^n - \zeta_{i-1}^n\right)\right]
$$

The inspection of Eq. (52) shows that $h_i^{n+1}$ varies quadratically with $\lambda$, and that $h_i^{n+1} \to h_i^n$ for $\lambda \to 0$. When $h_i^n > 0$, a continuity argument proves that there exists a sufficiently small time step $\Delta t > 0$ in the neighborhood of $\Delta t = 0$ that is able to ensure the condition $h_i^{n+1} \geq 0$ when $h_i^n > 0$. Troubles may arise when $h_i^n = 0$ and $(3\theta-1)\left(q_{i+1/2}^n - q_{i-1/2}^n\right) + (1-\theta)\left(q_{i+3/2}^n - q_{i-3/2}^n\right) > 0$, because $h_i^{n+1} < 0$ for every $\Delta t > 0$. These observations suggest a modification to the original algorithm where a limit depth value $\varepsilon_h$ is defined and the cells with $h_i^n > \varepsilon_h$ are flagged as wet, while the cells with $h_i^n \leq \varepsilon_h$ are flagged as dry. The algorithm is described as follows.

At each time level $t^n$, the algorithm iteratively reduces the time step, in order to find the value $\Delta t = \Delta t_j$ (where $\Delta t_j$ is the *j*-th guess) that satisfies the stability requirements and supplies positive depths $h_i^{n+1}$. At the beginning of the iterations, the initial guess $\Delta t = \Delta t_0$ is used, where the Eq. (48) is used for the calculation of $\Delta t_0$. At the interfaces between the cells, the position $q_{i-1/2}^{n+1} = 0$ is made when $h_{i-1}^n < \varepsilon_h$ and $h_i^n < \varepsilon_h$, otherwise the Eq. (46) is used to calculate the discharge $q_{i-1/2}^{n+1}$ between the cells $i$ and $i+1$. After this, the position $q_{i-1/2}^{n+1} = 0$ is made in the case that $q_{i-1/2}^{n+1} > 0$ and $h_{i-1}^n < \varepsilon_h$ or $q_{i-1/2}^{n+1} < 0$ and $h_i^n < \varepsilon_h$, because the mass flux must be limited in order to avoid the appearance of negative flow depth in dry cells. Finally, the flow depth in each cell is adjourned using Eq. (47). If $h_i^{n+1} < 0$ for some cell, the calculations are repeated with $\Delta t = \Delta t_j$, where $\Delta t_j = 0.7 \Delta t_{j-1}$, and the process is iterated until $h_i^{n+1} \geq 0$ everywhere.

In the following, $\varepsilon_h = 10^{-9}$ m is used.

*6.2.1 Riemann problems on flat bed (Tests 1 and 2)*

The modified *q-centred* scheme is applied with the same parameters to the Riemann problems discussed in the Subsection 6.1.1, and the corresponding results are represented in Figure 12. The inspection of the Figure 12a, where the numerical solution of Test 1 is represented at time $t = 5.001$ s, shows that there is no improvement of the numerical results with respect to the original *q-centred* scheme by de Almeida et al. (2012) for this problem. This enforces the conclusion that the error of the shock strength and position is due to the non-conservative nature of Eq. (46). The relative mass error is equal to $5.68 \cdot 10^{-16}$, which is very small as expected.

The numerical results of the Test 2 are represented for the time $t = 5.002$ in Figure 12b. The relative difference between the computed final volume and the theoretical final volume is $2.70 \cdot 10^{-15}$, and the comparison with the corresponding Riemann problem of Sub-section 6.1.1 shows that the modified algorithm improves the mass conservation property with respect to the original *q-centred* scheme. The inspection of Figure 12b shows that the modified algorithm exhibits a backward moving shock which moves towards left with celerity equal to the flow velocity, violating the Lax condition

(see Appendix C). This receding shock is a numerical artifact that arises because the numerical scheme supplies a calculation also in the case that the mathematical model has no exact solution.

[Insert Figure 12 about here]

*6.2.2 Oscillations in the parabolic channel (Test 3)*

The modified *q-centred* algorithm is applied to the Test 3, and the corresponding results are represented in Figure 13 for the times $t_1$ = 350.24 s (Figure 13a), $t_2$ = 710.52 s (Figure 13b), $t_3$ = 1060.79 s (Figure 13c), and $t_4$ = 1421.02 s (Figure 13d). The relative difference between the numerical and exact volume at time $t_4$ is $1.89 \cdot 10^{-10}$, showing a consistent improvement (nine orders of magnitude) with respect to the solution computed with the original *q-centred* scheme by de Almeida et al. (2012).

The comparison with the SWE exact results show that the new algorithm has improved capacity to capture the essentials of the solutions, namely the average slope of the free surface and the total mass. Nonetheless, the details of the representation remain unsatisfactory. For example, from Figure 13a it is evident that the algorithm creates to the left of the channel a spurious receding wet-dry front that moves from left to right, and this is in contrast with the structure of the LInA equations, which admit only advancing or standing wet-dry fronts. The advancing front on the right of Figure 13a is represented with a shock, congruently with the structure of the LInA model, but this is in contrast with the front continuously connected to the bed that is supplied by the SWE exact solution. In addition, the position of the wetting front does not coincide with that supplied by the SWE. During the deceleration of the flow, the left receding shock of Figure 13a is transformed in a smoothly varying wave that is visible at the centre of Figure 13b. When this wave reaches the right wet-dry front, the free surface exhibits a peaked shape to the right of Figure 13c. Successively, this peaked free surface profile develops as a receding shock that moves towards left, and this shock is in turn broken into the smoothly varying wave that is visible at the centre of Figure 13d. Finally, the maximum flow depth computed is $h_{max}$ = 3.62 m, and it is 13% greater than the exact value.

The reduction of the calculation time steps aiming at the satisfaction of a depth-positivity preserving property may lead to an unbearable increment of the computational time. For the present case, the minimum time step is $\Delta t_{min} = 6.32 \cdot 10^{-11}$ s, and it is $2.09 \cdot 10^{8}$ times smaller than the minimum time step used by the original algorithm. Such a small value of $\Delta t_{min}$ suggests that, depending on the dynamics of the moving wave and on the grid, the positivity of the algorithm cannot be easily ensured when receding fronts are present.

[Insert Figure 13 about here]

## 6.3 A Finite Volume scheme with Rusanov flux

In order to show the relation between the conservativity of the numerical scheme and the corresponding numerical results, a Rusanov Finite Volume scheme for the approximate solution of the LInA model is presented. The flow field is subdivided into rectangular cells of length $\Delta x$, and the significant variables $z_i$, $h_i$, $\zeta_i$, together with the discharge $q_i = h_i u_i$, are stored at the centre of the $i$-th cell. In each cell, the conserved variables $h_i$ and $q_i$ are advanced in time by means of the conservative scheme

(53) $\mathbf{u}_i^{n+1} = \mathbf{u}_i^n - \lambda \left[ \mathbf{f}_{i+1/2} - \mathbf{f}_{i-1/2} \right] + \lambda \left[ \mathbf{s}_{i-1/2}^+ + \mathbf{s}_{i+1/2}^- \right]$,

where the meaning of the symbols is as follows: $\mathbf{u}_i^n = \begin{pmatrix} h_i^n & q_i^n \end{pmatrix}^T$ is the vector of the conserved variables in the $i$-th cell; $T$ is the matrix transpose symbol; $\mathbf{f}_{i+1/2} = \begin{pmatrix} f_{h,i+1/2} & f_{q,i+1/2} \end{pmatrix}^T$ is the vector of the numerical fluxes $f_{h,i+1/2}$ (mass flux) and $f_{q,i+1/2}$ (momentum flux) through the interface $i+1/2$ between the cells $i$ and $i+1$; $\mathbf{s}^+_{i-1/2} = \begin{pmatrix} 0 & s^+_{i-1/2} \end{pmatrix}^T$ is the contribution of the interface $i-1/2$ to the source terms in the cell $i$-th; $\mathbf{s}^-_{i+1/2} = \begin{pmatrix} 0 & s^-_{i+1/2} \end{pmatrix}^T$ is the contribution of the interface $i+1/2$ to the source terms in the cell $i$-th.

Following the procedure by Audusse et al. (2014), the interface flow depths and the interface conserved quantities are first calculated by means

$$
\begin{aligned}
& z_{i+1/2} = \max\{z_i, z_{i+1}\} \\
(55) \quad & h^-_{i+1/2} = \max\{z_i + h_i - z_{i+1/2}, 0\}, \quad \mathbf{u}^-_{i+1/2} = \begin{pmatrix} h^-_{i+1/2} & q_i \end{pmatrix}^T \\
& h^+_{i+1/2} = \max\{z_{i+1} + h_{i+1} - z_{i+1/2}, 0\}, \quad \mathbf{u}^+_{i+1/2} = \begin{pmatrix} h^+_{i+1/2} & q_i \end{pmatrix}^T
\end{aligned}
$$

Successively, the Rusanov scheme (Toro 2009) is used to evaluate the numerical fluxes as

$$
(56) \quad \mathbf{f}_{i+1/2} = \frac{\mathbf{f}(\mathbf{u}^-_{i+1/2}) + \mathbf{f}(\mathbf{u}^+_{i+1/2})}{2} - c \frac{\mathbf{u}^+_{i+1/2} - \mathbf{u}^-_{i+1/2}}{2},
$$

where the maximum interface speed $c$ is calculated by means of $c = \sqrt{g \max\{h^-_{i+1/2}, h^+_{i+1/2}\}}$. Finally, the interface source terms are evaluated by means of

$$
(57) \quad \mathbf{s}^-_{i+1/2} = -0.5g \begin{pmatrix} 0 & (h_i)^2 - (h^-_{i+1/2})^2 \end{pmatrix}^T, \quad \mathbf{s}^+_{i+1/2} = 0.5g \begin{pmatrix} 0 & (h_{i+1})^2 - (h^+_{i+1/2})^2 \end{pmatrix}^T.
$$

The boundary conditions are imposed weakly, by adopting ghost cells outside of the computational domain (Toro 2001, 2009). For stability requirements, the time step satisfies the inequality $\Delta t \leq \Delta t_0$, where $\Delta t_0$ is evaluated by means of Eq. (48). In the following, the positions $\alpha = 0.5$ (Bouchut 2004), $\Delta t_{max} = 0.1$ s, and $\Delta t = \Delta t_0$, are constantly used.

*6.3.1 Riemann problems on flat bed (Tests 1 and 2)*

The Finite Volume scheme with Rusanov flux is applied to the two Riemann problems discussed in Subsection 6.1.1, and the corresponding results are represented in Figure 14. The numerical solution of the Test 1 is represented in Figure 14a, and compared with the exact solution. The inspection of the figure (solution at time $t = 5.002$ s) shows that the numerical solution is hardly distinguishable from the exact solution, because the Finite Volume scheme has the ability of capturing the moving shocks, due to its full conservativity. The relative mass error is modest, and equal to $2.13 \cdot 10^{-4}$, but it is not as small as the case of the *q-centred* schemes. Since no negative depth appears during the computations, the small mass conservation discrepancy is explained recalling that the boundary condition is enforced weakly in the Finite Volume scheme, and it is not imposed exactly as in the *q-centred* schemes.

The results for the Test 2 are represented for the time $t = 5.002$ in Figure 14b. Different from the *q-centred* simulations of Sub-section 6.1.1, the Finite Volume scheme exhibits a backward shock that moves towards left. The relative difference between the computed final volume and the theoretical final volume is 0.363, demonstrating that the receding shock generates a wake of cells with negative depth. It can be concluded that the adoption of a conservative scheme changes the wake rarefaction of Figure 10b into the shock of Figure 14b, but it does not solve the problem of the mass conservation when receding fronts are present. Nonetheless, the comparison with the corresponding result of Sub-section 6.1.1 shows that the fully conservative model has an improved mass-conservation property with respect to the *q-centred* scheme.

[Insert Figure 14 about here]

*6.3.2 Oscillations in the parabolic channel (Test 3)*

The Rusanov Finite Volume scheme is applied to the Test 3, and the corresponding results are represented in Figure 15 for the times $t_1$ = 350.20 s (Figure 15a), $t_2$ = 710.41 s (Figure 15b), $t_3$ = 1060.61 s (Figure 15c), and $t_4$ = 1420.80 s (Figure 15d). In the same figure, the exact solution of the SWE is also represented. The inspection of the plot shows that the Finite Volume scheme supplies a solution that is similar but not identical to the solution of the modified *q-centred* scheme, and it is quite different from the SWE exact solution.

The maximum value of the flow depth supplied by the numerical model during the calculations is $h_{max}$ = 3.73 m, and it is 16% greater than the exact value, These values show an improvement with respect to the original *q-centred* scheme, due to the increased mass-conservation ability of the scheme. This is confirmed by the relative difference between the numerical and exact volume at time $t_4$, which is equal to 0.204, and it is 53% less than the corresponding result for the original *q-centred* scheme.

[Insert Figure 15 about here]

### *6.4 A modified conservative Finite Volume scheme with Rusanov flux*

The Rusanov Finite Volume scheme is modified in order preserve the depth-positivity by reducing the time step, as made for the modified *q-centred* scheme. Again, a limit depth value $\varepsilon_h$ is defined, and the time step is iteratively reduced at each time level $t^n$ in order to find the value $\Delta t = \Delta t_j$ that supplies positive depths $h_i^{n+1}$. The Rusaonov flux is modified, and the position $\mathbf{f}_{i+1/2} = 0$ is made when one of the following conditions is true:

1) $f_{h,i+1/2} > 0$ and $h_{i+1/2}^- \leq \varepsilon_h$,

2) $f_{h,i+1/2} < 0$ and $h_{i+1/2}^+ \leq \varepsilon_h$,

3) $h_{i+1/2}^- \leq \varepsilon_h$ and $h_{i+1/2}^+ \leq \varepsilon_h$,

otherwise the fluxes are calculated as usual. If $h_i^{n+1} < 0$ for some cell, the calculations are repeated with $\Delta t = \Delta t_j$, where $\Delta t_j = 0.7 \Delta t_{j-1}$, and the process is iterated until $h_i^{n+1} \geq 0$ everywhere. In the following, $\varepsilon_h = 10^{-9}$ m is used.

*6.4.1 Riemann problems on flat bed (Tests 1 and 2)*

The modified Rusanov Finite Volume scheme is applied to the Test 1, and the numerical results are represented for t = 5.002 s in Figure 16a. The comparison with the exact Riemann problem solution shows that the modified Rusanov flux preserves the ability of capturing moving discontinuities. The relative mass error coincides with that of the original Rusanov Finite Volume scheme, confirming that this error is entirely due to the implementation of the boundary condition.

The results for the Test 2 are represented in Figure 16b with reference to the time $t$ = 5.001 s. The relative difference between the computed final volume and the theoretical final volume is significantly improved with respect to the original Rusanov scheme, and it is equal to $6.37 \cdot 10^{-4}$, but it is far from the machine epsilon, due to the implementation of the boundary condition. Finally, the comparison between Figures 14b and 16b shows that the improved mass-conservation property of the modified Rusanov scheme significantly shifts the position of the receding shock.

[Insert Figure 16 about here]

*6.4.2 Oscillations in the parabolic channel (Test 3)*

The modified Rusanov Finite Volume scheme is applied to the Test 3, and the numerical results corresponding to the times $t_1$ = 350.22 s and $t_2$ = 710.41 s are represented in Figure 17. The relative difference between the numerical and exact volume at time $t_2$ is $1.89 \cdot 10^{-10}$, demonstrating that the mechanism of iterative time-step reduction may contribute to the mass conservation during the calculations. Nonetheless, it must be observed that the last saved result is at time $t_{last}$ = 930.51 s, because the computer program enters an infinite loop. This loop is caused by the fact that the computational scheme does not find a time step that ensures the positivity of the results during the entire duration of calculations.

[Insert Figure 17 about here]

# 7 Discussion

The results presented in the preceding sections are discussed for their implications on flooding modelling with the LInA model. In particular, the differences between LInA and SWE are commented, and novel applicability limits for the LInA model are discussed. Finally, the numerical modelling of the LInA model is critically reviewed, and the mathematical results demonstrated in the preceding sections are extended to the two-dimensional case.

## *7.1 Moving frontier modelling*

In the LInA model, the flooding phenomenon is characterized by an unphysical feature described in Sections 2 and 3, namely the presence of a shock at the wetting frontier. The numerical experiments available in the literature have never evidenced this frontal shock, probably because the numerical diffusion effects are able to smoothen discontinuous solutions, and because the examination of large flooded area maps is not informative with the reference to the presence of such a steep frontier.

The speed of the LInA frontal shock is minor than the speed of the fastest signal in the SWE, as confirmed by the comparison between the solutions of the Riemann problem on dry bed supplied by LInA and SWE (see Figures 3 and 4). Commonly, the speed of the wetting front is used to calibrate the floodplain roughness, and this implies that the slower front propagation speed should be compensated by lower values of the roughness (de Almeida and Bates 2013). Nonetheless, this could lead to an underestimation of the roughness in the flooded areas far from the wetting front.

In order to ensure the physical representativeness of the flooding simulation, the limitation of the shock height at the wetting front is desirable. It is clear from Eqs. (19) and (20) that the flow depth $h_F$ and velocity $u_F$ at the wetting front are connected by $|u_F| = \sqrt{gh_F/2}$, and this means that a control on the velocity of the wetting front could be used for rejection or acceptance of flooding simulation. For example, if the acceptable shock height at the wetting front is $h_{\lim} = 0.02$ m, the limit velocity is $u_{\lim} = 0.31$ m/s, and the simulations with $u_F > u_{\lim}$ should be rejected. Lower values of the shock height at the wetting front could constitute a very severe requirement for the definition of the cases where the LInA model is acceptable for flooding simulations. Conversely, higher values of the shock height could relax this flow velocity limit, but this could be unacceptable for the physical soundness of results.

A significant result found in Section 3 is that the drying of the wet bed is forbidden. This is a big issue, because receding flows are likely to happen during floods caused by rainfalls with variable intensity or multipeaked hydrographs, meaning that many practical applications cannot be modeled by means of the LInA model. Practitioners should be aware that the LInA numerical solutions with receding fronts are merely an algorithmic creation with no physical counterpart.

*7.2 Effects of uneven bed elevation and obstacles*

In the Section 3, the impact of a wave against an obstacle has been studied. The corresponding analytic results show that the LInA model underestimates the flow depth at the obstacle, while the impact calms the flow down more rapidly than the SWE. The first consequence is that the flow velocity in the vicinity of obstacles may be underestimated, with consequences on the evaluation of road infrastructures damaging (Kreibich et al. 2011) and pedestrian stability (Arrighi et al. 2017). Similarly, the underestimation of the flow depth may have consequences for the evaluation of pedestrian stability (Arrighi et al. 2017), for the safety of transportation (Pregnolato et al. 2017), and for the evaluation of building damage (Pistrika et al. 2014). It is evident that these observations especially apply with reference to the evaluation of flooding damages and human safety in urban areas, where these aspects are prominent for the presence of numerous buildings and obstacles with complicate geometry.

Another consequence of flow depth underestimation at obstacles is that there are conditions in which the SWE model exhibits sufficient energy to make the flow jump over the obstacle, while the LInA predicts full reflection (see the Riemann problem of Figure 7b). It is clear that the inaccurate treatment of obstacles may exclude large portions of land from the computation of the flooded areas, and that the flooded area error increases with the flooding duration.

The example Riemann problems in Section 5 show that the flow energy at bed elevation discontinuities is not conserved by the LInA model. In particular, energy is acquired at bed steps (positive bed elevation discontinuities) while energy is loss at bed drops (negative bed elevation discontinuities). In order to shed light on this fact, the conservation of energy for the LInA model is studied. Simple manipulations of Eq. (2) allow writing the following equation

$$(58) \quad \frac{\partial E_m(\mathbf{U})}{\partial t} + \frac{\partial}{\partial x}\left[ghu H(\mathbf{U})\right] = u\frac{\partial hu^2}{\partial x} - ghuS_f,$$

which represents the balance equation for the mechanical energy $E_m(\mathbf{U}) = gh\left(h/2 + u^2/2 + z\right)$. From Eq. (58) it is evident that the mechanical energy is not conserved, even in the case that friction is neglected, because the source term $u\partial hu^2/\partial x$ is present. If the friction is negligible, the steady flow solution of Eq. (2) reduces to

$$\text{(59)} \quad \begin{aligned} &\frac{\partial hu}{\partial x} = 0 \\ &\frac{\partial}{\partial x} H(\mathbf{U}) = \frac{\partial}{\partial x} \frac{u^2}{2g} \end{aligned}.$$

The Eq. (59) states that the head is not conserved through the domain, and this is in contrast with what is expected from physical intuition. After some algebraic manipulation, the integration of Eq. (59) between the abscissas $x_1$ and $x_2$ (with $x_1 < x_2$) leads to:

$$\text{(60)} \quad \frac{H(\mathbf{U}_1) - H(\mathbf{U}_2)}{h_1} = \frac{F^2(\mathbf{u}_1)}{2} \frac{\beta^2 - 2\beta}{(1-\beta)^2},$$

where $\mathbf{u}_i = \begin{pmatrix} h_i & h_i u_i \end{pmatrix}^T$ and $\mathbf{U}_i = \begin{pmatrix} h_i & h_i u_i & z_i \end{pmatrix}^T$ are the vector of the conserved variables and the augmented vector of the conserved variables at the abscissa $x_i$ ($i$ = 1, 2), respectively, while $\beta = (z_2 - z_1)/h_1$ is the relative bed elevation variation.

This analysis confirms the results contained in Section 5 for the Riemann problem at bed steps and drops, because the Eq. (60) clarifies that head is gained if the bed slope is negative ($z_1 > z_2$), while head is lost in the case of positive bed slope ($z_1 < z_2$). The gain of energy with bed elevation variations is clearly in contrast with what is expected from physics, and this allows the introduction of a limit to acceptable Froude numbers, in order to save the physical soundness of results. For example, if the acceptable value for the spurious relative head variation is $(H(\mathbf{U}_1) - H(\mathbf{U}_2))/h_1 = -0.05$, the Froude number $F(\mathbf{u}_1)$ must be smaller than 0.21 for $\beta = 0.45$.

The study of LInA exact solutions has evidenced an additional result, namely the fact that there are initial conditions for which the Riemann problem at the bed drop has no solution. This finding is of the greatest importance for flood modelling, because terrain steps and drops are commonly present in real world applications, as discussed in Section 5. In addition, topographic data are usually supplied as a rectangular two-dimensional grid where the terrain elevation is constant in each cell, and this implies that bed elevation drops are always present at the interface between cells in the original representation of the input data. Of course, the numerical applications available in the literature have never detected such an issue of the LInA model because the bed drops between computational cells are usually treated as linear ramps whose slope becomes infinity only in the case of mesh refinement.

The inspection of Eq. (44) demonstrates that the non-existence of the LInA solution at bed drops is not related to an exotic class of initial conditions. In order to show this, it is sufficient to recall that the wetting fronts of the LInA model are shocks that move on the dry bed with Froude number equal to $F_w$. In this case, the position $F(\mathbf{u}_L) = F_w$ in Eq. (44) leads to the limit value $a/h_L = 1.56$ of the relative drop height. For example, with $h_F$ = 0.01 m and $u_F$ = 0.22 m/s, the maximum admissible drop height is $a_{max}$ = 0.016 m. This result is particularly discouraging, because $a_{max}$ = 0.016 m is far smaller than the height of the typical bed elevation irregularities that are found in flooding problems. In these circumstances, it is clear that the approximate solutions supplied by numerical models have no meaning, because the analytic solution of the flood propagation problem does not exist. This poses a severe conceptual limit to the use of the mathematical model.

*7.3 Numerical modelling*

In Section 6, two different numerical models have been considered, namely the *q-centred* scheme by de Almeida et al. (2012), which is not fully conservative because of the rearrangement of the momentum equation, and a novel fully conservative Rusanov Finite Volume scheme.

The numerical experiments have confirmed the mathematical analysis of Section 2, showing that a shock is spontaneously generated at wetting fronts in all the computations. The presence of a shock on the wetting front implies that numerical models written in non-conservative form cannot be used for flooding propagation (Hou and LeFloch 1994), because they introduce an unavoidable error in the computation of shock speed and strength. This error could in turn adversely affect the roughness coefficient calibration. The inspection of the results supplied for the Test 1 by the *q-centred* scheme (Sub-section 6.1.1) and by the Rusanov Finite Volume scheme (Sub-section 6.3.1) confirm the theoretical finding by Hou and LeFloch (1994).

Interestingly, the Test 2 and the Test 3 show that different numerical models produce very different solutions at the receding fronts, confirming the lack of reliability of the corresponding numerical results. For example, the *q-centred* scheme produces receding fronts characterized by a sort of rarefaction wave (see Figure 10b for Test 2 and Figure 11 for Test 3), while the Rusanov Finite Volume scheme produces a receding shock (see Figure 14b for Test 2 and Figure 15 for Test 3). Of course, the theoretical analysis shows that the free surface cannot be continuously connected to the bed, excluding the correctness of the receding rarefaction in the *q-centred* scheme, while the structure of the admissible shocks excludes the correctness of the receding shock in the Rusanov Finite Volume scheme. These observations demonstrate that the LInA numerical models are able to supply a solution, which is merely an algorithmic creation, also in the case of receding fronts, and may explain why the non-existence of receding fronts has never been individuated in the past literature.

The numerical experiments of Sub-sections 6.1 and 6.3 have shown that a major source of numerical error in the original *q-centred* scheme and in the Rusanov Finite Volume scheme is the lack of volume conservation in the cells where the wet bed dries. The corresponding modified algorithms proposed in Sub-sections 6.2 and 6.4 are based on two ingredients, namely the reduction of the time step and the flux limitation. The reduction of the time step is a standard procedure that enhances algorithms stability by avoiding that the flow depth becomes negative in wet cells (Burguete et al. 2007), and it has no influence on the structure of the problem solution. Conversely, the flux limitation, which is required to avoid that discharge flows out from dry cells, mutates the receding rarefactions of the *q-centred* scheme in receding shocks that are forbidden by the LInA model mathematical structure and that violate the Lax entropic condition.

Unfortunately, the modified algorithms introduce an adverse effect, namely the unacceptable increase of computational burden caused by the uncontrolled decrease of the time step (as evidenced by Test 3 in Sub-Section 6.3.2), and possibly the infinite loop of the algorithm (Test 3 of Sub-section 6.4.2). It is evident that the time-step reduction strategy, which works nicely for the SWE conservative numerical schemes, is not helpful in the case of the LInA schemes because the LInA mathematical model does not support exact solutions with receding flow. In other words, a well-written LInA numerical model cannot supply better results than the LInA mathematical model.

*7.4 Two-dimensional modelling*

The *Proposition* 1 and the *Corollary* 1 have been demonstrated with reference to the one-dimensional LInA model. Nonetheless, it is easy to show that the same results can be extended without special restrictions to the full two-dimensional case, as follows. The two-dimensional LInA equations can be written as (Aronica et al. 1998, Moramarco et al. 2005):

$$(61) \quad \begin{aligned} &\frac{\partial h}{\partial t} + \frac{\partial hu}{\partial x} + \frac{\partial hv}{\partial y} = 0 \\ &\frac{\partial hu}{\partial t} + \frac{\partial}{\partial x}\frac{gh^2}{2} = -gh\frac{\partial z}{\partial x} - ghS_{f,x}, \\ &\frac{\partial hv}{\partial t} + \frac{\partial}{\partial y}\frac{gh^2}{2} = -gh\frac{\partial z}{\partial y} - ghS_{f,y} \end{aligned}$$

where $y$ is the horizontal coordinate normal to $x$ and $v$ the corresponding component of the velocity, while $S_{f,x}$ and $S_{f,y}$ are the components along $x$ and $y$ of the friction slope, respectively. If the flow is smooth, the system of Eq. (61) can be rewritten as

$$\frac{\partial h}{\partial t} + h\left(\frac{\partial u}{\partial x} + \frac{\partial v}{\partial y}\right) + u\frac{\partial h}{\partial x} + v\frac{\partial h}{\partial y} = 0$$

(62) $\quad h\dfrac{\partial u}{\partial t} + u\dfrac{\partial h}{\partial t} + gh\dfrac{\partial h}{\partial x} = -gh\dfrac{\partial z}{\partial x} - ghS_{f,x}$,

$$h\frac{\partial v}{\partial t} + v\frac{\partial h}{\partial t} + gh\frac{\partial h}{\partial y} = -gh\frac{\partial z}{\partial y} - ghS_{f,y}$$

and it reduces to

$$\frac{\partial h}{\partial t} + u\frac{\partial h}{\partial x} + v\frac{\partial h}{\partial y} = 0$$

(63) $\quad u\dfrac{\partial h}{\partial t} = 0$

$$v\frac{\partial h}{\partial t} = 0$$

at the wet-dry frontier, where $h \to 0$. The first of Eq. (63) states that the material derivative $dh/dt = \partial h/\partial t + u\,\partial h/\partial x + v\,\partial h/\partial y$ of the flow depth on the wet-dry frontier with $h = 0$ is null, and this implies that the fluid particle on the wet-dry frontier remains attached to this frontier, as expected. The second and the third of Eq. (63) are contemporarily verified in the cases (a) $u = 0$ and $v = 0$ or (b) $\partial h/\partial t = 0$. The case (a) states that the flow velocity is null at the frontier, implying that the frontier itself is fixed. The case (b) states that the variation in time of flow depth at the position characterized by $h = 0$ is null, implying again that the frontier is fixed. In other words, the *Proposition* 1 of Section 2 is valid also in the two-dimensional case, and the surface profile can be continuously connected to the dry bed only in the case of fixed frontier. By exclusion, the moving wet-dry frontiers of the two-dimensional LInA model can be moving shocks only, confirming that the *Corollary* 1 of Section 2 is valid also in the two-dimensional case.

In order to extend the theory developed in the Section 3 to the two-dimensional LInA, it is sufficient to demonstrate that the two-dimensional shocks behave locally as the one-dimensional shocks. In the case of the two-dimensional LInA model, the Rankine-Hugoniot condition can be written as (Dafermos 2005)

$$(64) \quad \left[\mathbf{G}(\mathbf{U}) - \mathbf{G}(\mathbf{U}_0)\right] \cdot \mathbf{n} = \sigma(\mathbf{U} - \mathbf{U}_0),$$

where $\mathbf{n} = \begin{pmatrix} n_x & n_y \end{pmatrix}^T$ is the unit vector that is locally normal to the shock front, $\sigma$ is the celerity of the shock, $\mathbf{U} = \begin{pmatrix} h & hu & hv \end{pmatrix}^T$ is the state to the right of the moving shock and $\mathbf{U}_0 = \begin{pmatrix} h_0 & h_0 u_0 & h_0 v_0 \end{pmatrix}^T$ is the state to the left, while the flux matrix $\mathbf{G}$ is defined as

$$(65) \quad \mathbf{G}(\mathbf{U}) = \begin{pmatrix} hu & hv \\ 0.5gh^2 & 0 \\ 0 & 0.5gh^2 \end{pmatrix}.$$

The vectorial Eq. (64) can be expressed in scalar form as

$$(66) \quad \begin{aligned} (hu - h_0 u_0) n_x + (hv - h_0 v_0) n_y &= \sigma(h - h_0) \\ \frac{g}{2}(h^2 - h_0^2) n_x &= \sigma(hu - h_0 u_0) \\ \frac{g}{2}(h^2 - h_0^2) n_y &= \sigma(hv - h_0 v_0) \end{aligned},$$

and some algebra allows rewriting

$$(67) \quad \begin{aligned} hU - h_0 U_0 &= \sigma(h - h_0) \\ \frac{g}{2}(h^2 - h_0^2) &= \sigma(hU - h_0 U_0), \\ hV - h_0 V_0 &= 0 \end{aligned}$$

where $U = un_x + vn_y$ is the component of the flow velocity that is normal to the shock, while $V = -un_y + vn_x$ is the tangential component. Notably, the first two equations of the system (67) can be solved independently from the third equation. Most important, these two equations reduce to the one-dimensional shock definition (see Eq. [C.1] of Appendix C) along the axis that is normal to the shock front. For this reason, the *Propositions* 2, 3, and 4, which depend on the one-dimensional shock definition, remain valid in the two-dimensional case.

Finally, the *Proposition* 5 states that there is a class of initial conditions for which the one-dimensional LInA model has no solution. Of course this result is immediately generalized to the two-dimensional case by recalling that the one-dimensional model is the particular case of the two-dimensional model where the derivatives and the velocity components along *y* are null.

In conclusion, the two-dimensional LInA model admits only solutions with increase of the wetted area, while the receding of the wet-dry frontiers is forbidden, and it is possible to consider a class of initial conditions for which the two-dimensional LInA model has no solution.

## 8 Conclusions

Despite the fact that the Local Inertia Approximation (LInA) has been applied for years in the field of flooding, the researchers have dedicated scarce efforts to a closer examination of its physical justification and of the corresponding exact solutions. The study of the Shallow water Equations (SWE) in dimensionless form shows that, far from the wetting front, the LInA model is not more accurate than the Noninerta Approximation (NIA), and it is even a worse approximation at the wetting front. The study of moving wet-dry frontiers in the LInA model has shown that these are always characterized by a shock, and this prompts the general study of moving discontinuities. For this reason, the complete solution of the Riemann problem has been tackled, demonstrating that receding shocks are forbidden. The consequence is that the drying of the wet bed in the LInA model is forbidden. In addition, the study of the nonbreaking wave on horizontal bed with friction has demonstrated that the formation of the shock at the wetting front is not confined to the frictionless case, but it is a completely general phenomenon for the LInA model.

Bed elevation discontinuities are common features of natural geometries and urban terrain configurations. The study of the Riemann problem with discontinuous bed elevation has demonstrated that the LInA model exhibits unphysical variation of energy at bed discontinuities, and that an underestimation of the flood stage and flooded areas is possible when obstacles (bumps, levees, sidewalk steps, walls) are present. In addition, there are cases where the presence of a bed drop may cause the crisis of the mathematical model because no exact solution is possible. Actually, the non-existence of the LInA solution in the case of discontinuous topography and the non-existence of receding fronts radically question the viability of the LInA model in realistic cases.

The numerical analysis has considered two different numerical models, namely the non-conservative *q-centred* algorithm by de Almeida et al. (2012) and the fully conservative Rusanov Finite Volume scheme. The numerical tests have demonstrated that the LInA numerical models are able to produce numerical solutions (for example receding flows) also in the cases that the mathematical solution does not exist. In the case of drying fronts, the numerical solutions are characterized by the lack of mass conservation. The attempt of reducing the mass unbalance by reducing the time step and by limiting the mass fluxes is an unsatisfactory strategy, because the computational burden increases enormously, and possibly leads to infinite loops. Of course, the numerical solutions without mathematical counterpart, such as those with receding fronts and irregular topography, should be simply rejected.

Based on the preceding theoretical results, the definition of two applicability limits for the LInA model has been examined, limited to the unusual case of continuously varying bed elevation with absence of receding fronts. The first criterion refers to the velocity of the wetting front, and it is established in order to limit the height of the unphysical shock at the wet-dry frontier. The second criterion is based on the limitation of spurious energy variations through the domain. These criteria show that the applicability limits of the LInA model are discouragingly severe, even if the bed elevation varies continuously. Of course, no applicability is possible in the case of receding fronts, or in the case of bed elevation drops whose height is 56% greater than the height of the frontal shock. From these considerations, it is evident that classic SWE models should be preferred in the majority of the practical applications.

**Acknowledgments**

Dr. Eng. Gabriella Petaccia is gratefully acknowledged for having supplied a postprint copy of the paper by Luigi Natale and Fabrizio Savi that is cited in the References list.

This research was partially funded by the University of Naples Parthenope through the funding program "Sostegno alla Ricerca Individuale 2015-2017" and "Ricerca competitiva triennio 2016-2018".

**Appendices**

*A. Genuine non-linearity of the 1- and 2-waves in the system of Eq. (13)*

In the phase plane ($h$, $hu$), the integral curves of the first characteristic field corresponding to the system of Eq, (13) are the solutions of the following ordinary differential equation (LeVeque 1992)

(A.1) $\dfrac{d\mathbf{u}}{ds} = \alpha(s)\mathbf{r}_1(\mathbf{u}(s))$,

where $\alpha(s) \neq 0$ is a scaling factor. It is easy to see that the following is true along these curves:

(A.2) $\dfrac{d\lambda_1}{ds} = \nabla^T \lambda_1(\mathbf{u}(s))\dfrac{d\mathbf{u}}{ds} = \alpha(s)\nabla^T \lambda_1(\mathbf{u}(s))\mathbf{r}_1(\mathbf{u}(s)) = -\alpha(s)\dfrac{1}{2}\sqrt{\dfrac{g}{h}} \neq 0.$

This implies that the first characteristic field is genuinely non-linear, because $\lambda_1$ varies monotonically along the corresponding integral curves (LeVeque 1992). In a similar manner, it can be demonstrated that the second characteristic field is genuinely non-linear.

*B. Rarefaction waves for the system of Eq. (13)*

Eliminating $\alpha(s)ds$ from Eq. (A.1), the following

(B.1) $\dfrac{d\,h}{1} = \dfrac{d\,hu}{-\sqrt{gh}}$

is valid along the integral curves of the first characteristic field. The Eq. (B.1) can be solved with initial condition $\mathbf{u}(0) = \mathbf{u}_0$, supplying the explicit expression of the integral curve:

(B.2) $hu + 2h\sqrt{gh}/3 = h_0 u_0 + 2h_0\sqrt{gh_0}/3.$

The Eq. (B.2) is valid for $h > 0$, because the velocity $u$ is infinite for $h = 0$. By definition (LeVeque 1992), the right state $\mathbf{u}$ is connected to the left state $\mathbf{u}_0$ by a direct rarefaction wave contained into the first characteristic field if $\mathbf{u}$ lies on the integral curve of Eq. (B.2), and the condition $\lambda_1(\mathbf{u}_0) < \lambda_1(\mathbf{u})$ is satisfied. It follows the definition

(B.3) $R_1(\mathbf{u},\mathbf{u}_0): \quad 0 < h \leq h_0, \quad hu = h_0 u_0 + 2\left(h_0\sqrt{gh_0} - h\sqrt{gh}\right)/3.$

A similar reasoning, which is not reported here for the sake of brevity, shows that the right state $\mathbf{u}_0$ is connected to the left state $\mathbf{u}$ by a backward rarefaction wave contained in the second characteristic field if $\mathbf{u}$ lies on the integral curve of the second characteristic field, and the condition $\lambda_2(\mathbf{u}) < \lambda_2(\mathbf{u}_0)$ is satisfied. It follows the definition

(B.4) $R_2^B(\mathbf{u},\mathbf{u}_0): \quad 0 < h \leq h_0, \quad hu = h_0 u_0 - 2\left(h_0 \sqrt{gh_0} - h\sqrt{gh}\right)/3$.

## C. Shock waves for the system of Eq. (13)

If the constant states $\mathbf{u}$ and $\mathbf{u}_0$ are separated by a discontinuity that is propagating with speed $\sigma$, the Rankine-Hugoniot condition must be satisfied (LeVeque 1992). For the system of Eq. (13), this can be written as:

(C.1) $\begin{aligned} hu - h_0 u_0 &= \sigma(h - h_0) \\ \frac{gh^2}{2} - \frac{gh_0^2}{2} &= \sigma(hu - h_0 u_0) \end{aligned}$.

Some algebra shows that, for a given state $\mathbf{u}_0$, the Hugoniot locus of the states $\mathbf{u}$ that satisfy the Eq. (C.1) is represented in the phase plane $(h, hu)$ by

(C.2) $hu = h_0 u_0 \pm (h - h_0)\sqrt{0.5g(h + h_0)}$,

with corresponding shock speed

(C.3) $\sigma = \pm\sqrt{0.5g(h + h_0)}$.

In Eqs. (C.2)-(C.3), the sign minus and the sign plus define two distinct Hugoniot curves that are tangent to the eigenvectors $\mathbf{r}_1(\mathbf{u}_0)$ and $\mathbf{r}_2(\mathbf{u}_0)$, respectively. By definition (LeVeque 1992), the right state $\mathbf{u}$ is connected to left state $\mathbf{u}_0$ the by a shock wave contained into the first characteristic field if $\mathbf{u}$ lies on the Hugoniot curve tangent to $\mathbf{r}_1(\mathbf{u}_0)$, and the Lax condition $\lambda_1(\mathbf{u}_0) > \sigma > \lambda_1(\mathbf{u})$ is satisfied. It follows the definition

(C.4) $S_1(\mathbf{u},\mathbf{u}_0)$: $h \geq h_0$, $hu = h_0 u_0 + \sigma_1(\mathbf{u},\mathbf{u}_0)(h - h_0)$, $\sigma_1(\mathbf{u},\mathbf{u}_0) = -\sqrt{0.5g(h+h_0)}$.

Similarly, the left state $\mathbf{u}$ is connected to the right state $\mathbf{u}_0$ by a backward shock contained into the first characteristic field if $\mathbf{u}$ lies on the Hugoniot curve tangent to $\mathbf{r}_2(\mathbf{u}_0)$, and the Lax condition $\lambda_2(\mathbf{u}) > \sigma > \lambda_2(\mathbf{u}_0)$ is satisfied. It follows the definition

(C.5) $S_2^B(\mathbf{u},\mathbf{u}_0)$: $h \geq h_0$, $hu = h_0 u_0 + \sigma_2(\mathbf{u},\mathbf{u}_0)(h - h_0)$, $\sigma_2(\mathbf{u},\mathbf{u}_0) = \sqrt{0.5g(h+h_0)}$.

### D. Monotonicity properties of the wave curves for the system of Eq. (13)

It is easy to demonstrate that the direct 1-wave curve $H_1$ of Eq. (16) is a continuous, strictly decreasing, and strictly concave function in the plane $(h, hu)$. In particular, it is possible to show that the first and the second derivative of $hu$ with respect to $h$ are negative on both the rarefaction and the shock part of the wave curve:

(D.1)
$$0 < h < h_0, \quad \frac{d\,hu}{dh} = -\sqrt{gh} < 0, \quad \frac{d^2 hu}{dh^2} = -\frac{1}{2}\sqrt{\frac{g}{h}} < 0$$
$$h > h_0, \quad \frac{d\,hu}{dh} = -\frac{3gh + gh_0}{\sqrt{8g(h+h_0)}} < 0, \quad \frac{d^2 hu}{dh^2} = -\frac{12g^2 h + 20g^2 h_0}{\left[8g(h+h_0)\right]^{3/2}} < 0$$

The continuity in $h_0$ of $hu$ and of the corresponding derivatives follows immediately from the expressions of Eq. (16) and Eq. (D.1). Similarly, it is possible to show that the backward 2-wave $H_2^B$ of Eq. (17) is a continuous, strictly increasing, and strictly convex function in the plane $(h, hu)$.

*E. Contact discontinuities*

The elimination of $\alpha(x)dx$ from Eq. (39) supplies

(E.1) $\dfrac{dh}{1} = \dfrac{dz}{-1}, \quad d\,hu = 0,$

which leads to the solution

(E.2) $z + h = const., \quad hu = const.$

**References**

Abbott M.B. (1966) An introduction to the method of characteristics, Thames and Hudson, London.

Alcrudo F., Benkhaldoun F. (2001) Exact solutions to the Riemann problem of the shallow water equations with a bottom step, Computers & Fluids 30(6), 643-671. Doi: 10.1016/S0045-7930(01)00013-5.

Aronica T., Tucciarelli T., Nasello C. (1998) 2D Multilevel model for flood wave propagation in flood-affected areas, ASCE Journal of Hydraulic Engineering 124(4), 210-217. Doi: 10.1061/(ASCE)0733-9496(1998)124:4(210).

Audusse E., Bouchut F., Bristeau M.-O., Klein R., Perthame B. (2004) A fast and stable well-balanced scheme with the hydrostatic reconstruction for shallow water flows. SIAM Journal of Scientific Computing 25(6), 2050-2065. Doi: 10.1137/S1064827503431090.

Arrighi C., Oumeraci H., Castelli F. (2017) Hydrodynamics of pedestrians' instability in floodwaters, 21(1) 515-531. Doi: 10.5194/hess-21-515-2017.

Bates P.D., Horritt M.S., Fewtrell T.J. (2010) A simple inertial formulation of the shallow water equations for efficient two-dimensional flood inundation modelling, Journal of Hydrology, 387(1-2), 33-45. Doi: 10.1016/j.jhydrol.2010.03.027.

Bouchut F. (2004) Nonlinear stability of finite volume methods for hyperbolic conservation laws and well-balanced schemes for sources. Birkhäuser Verlag, Basel.

Burguete J., García-Navarro P., Murillo J., García-Palacín I. (2007) Analysis of the friction term in the one-dimensional Shallow-water model, ASCE Journal of Hydraulic Engineering 133(9), 1048-1063. Doi: 10.1061/(ASCE)0733-9429(2007)133:9(1048).

Cea L., Bladé E. (2015) A simple and efficient unstructured finite volume scheme for solving the shallow water equations in overland flow applications, Water Resources Research 51(7), 5464-5486. Doi: 10.1002/2014WR016547.

Chanson H. (2005) Analytical solution of dam break wave with flow resistance. Application to Tsunami surges. In: Jun B.H., Lee S.I., Seo I.W., Choi G.W. (eds.) Proceedings of the 31st IAHR Biennial Congress, Korea Water Resources Association, Seoul.

Chen Y., Zhou H., Zhang H., Du G., Zhou J. (2015) Urban flood risk warning under rapid urbanization, Environmental Research 139, 3-10. Doi: 10.1016/j.envres.2015.02.028.

Coulthard T.J., Neal J.C., Bates P.D., Ramirez J., de Almeida G.A.M., Hancock G.R. (2013) Integrating the LISFLOOD-FP 2D hydrodynamic model with the CAESAR model: implication for modelling landscape and evolution, Earth Surface Processes and Landforms 38(15), 1897-1906. Doi: 10.1002/esp.3478.

Cozzolino L., Della Morte R., Covelli C., Del Giudice G., Pianese D. (2011) Numerical solution of the discontinuous-bottom Shallow-water Equations with hydrostatic pressure distribution at the step, Advances in Water Resources 34(11), 1413-1426. Doi: 10.1016/j.advwatres.2011.07.009.

Cozzolino L., Pepe V., Morlando F., Cimorelli L., D'Aniello A., Della Morte R., Pianese D. (2017) Exact solution of the dam-break problem for constrictions and obstructions in constant width rectangular channels, ASCE Journal of Hydraulic Engineering 143(11), 04017047. doi: 10.1061/(ASCE)HY.1943-7900.0001368.

Cunge J.A., Holly F.M., Verwey A. (1980) Practical aspects of computational river hydraulics. Pitman, Boston.

Dafermos C.M. (2005) Hyperbolic Conservation Laws in Continuum Physics. Springer Verlag, Berlin.

de Almeida G.A.M., Bates P. (2013) Applicability of the local inertial approximation of the shallow water equations to flood modeling, Water Resources Research 49(8), 4833-4844. Doi: 10.1002/wrcr.20366.

de Almeida G.A.M., Bates P., Freer J.E., Souvignet M. (2012) Improving the stability of a simple formulation of the shallow water equations for 2-D flood modeling, Water Resources Research 48(5), W05528. Doi: 10.1029/2011WR011570.

Dingman S.L. (2009) Fluvial hydraulics, Oxford University Press, Oxford.

Doocy S., Daniels A., Murray S., Kirsch T.D. (2013) The Human Impact of Floods: a Historical Review of Events 1980-2009 and Systematic Literature Review, PLOS Current disasters. Doi: 10.1371/currents.dis.f4deb457904936b07c09daa98ee8171a.

Dottori F., Todini E. (2011) Developments of a flood inundation model based on the cellular automata approach: Testing different methods to improve model performance, Physics and Chemistry of the Earth 36(7-8), 266-280. Doi: 10.1016/j.pce.2011.02.004.

Dressler R.F. (1952) Hydraulic resistance effect upon dam-break functions, Journal of Research of the National Boureau of Standards 49(3), 217-225.

Duran A., Marche F., Turpault R., Berthon C. (2015) Asymptotic preserving scheme for the shallow water equations with source terms on unstructured meshes, Journal of Computational Physics 287, 184-206. Doi: 10.1016/j.jcp.2015.02.007.

Eymard R., Gallouët T., Herbin R., Latché J.-C. (2007) Analysis tools for Finite Volume schemes, Acta Mathematica Universitatis Comenianae 76(1), 111-136. http://www.iam.fmph.uniba.sk/amuc/_vol-76/_no_1/_herbin/herbin.pdf

Falter D., Vorogushyn S., Lhomme J., Apel H., Gouldby B., Merz B. (2013) Hydraulic model evaluation for large-scale flood risk assessments, Hydrological Processes 27(9), 1331-1340. Doi: 10.1002/hyp.9553.

Fowler A. (2011) Mathematical Geoscience, Springer-Verlag, London.

Galbiati G., Savi F. B. (1995) Evaluation of the comparative influence of soil hydraulic properties and roughness on overland flow at the local scale, Journal of Agriculture Engineering Research 61(3), 183-190. Doi: 10.1006/jaer.1995.1045.

Hou T.Y, LeFloch P.G. (1994) Why nonconservative schemes converge to wrong solutions: error analysis, Mathematics of Computation 62, 497-530. Doi: 10.1090/S0025-5718-1994-1201068-0.

Hunter N.H., Horritt M.S., Bates P.D., Wilson M.D., Werner M.G.F. (2005) An adaptive time step solution for raster-based storage cell modelling of floodplain inundation, Advances in Water Resources 28(9), 975-991. Doi: 10.1016/j.advwatres.2005.03.007.

Hunter N.H., P.D. Bates, M.S. Horritt, M.D. Wilson (2007) Simple spatially-distributed models for predicting flood inundation: a review, Geomorphology 90(3-4), 208-225. Doi: 10.1016/j.geomorph.2006.10.021.

Kreibich H., Piroth K., Seifert I., Maiwald H., Kunert U., Schwarz J., Merz B., Thieken A.H. (2009) Is flow velocity a significant parameter in flood damage modelling?, Natural Hazard and Earth System Sciences 9(5), 1679-1692. Doi: 10.5194/nhess-9-1679-2009.

Kundzewicz Z.W., Kanae S., Seneviratne S.I., Handmer J., Nicholls N., Peduzzi P., Mechler R., Bouwer L.M., Arnell N., Mach K., Muir-Wood R., Brakenridge G.R., Kron W., Benito G., Honda Y., Takahashi K., Sherstyukov B. (2014) Flood risk and climate change: global and regional perspectives, Hydrological Sciences Journal 59(1), 1-28. Doi: 10.1080/02626667.2013.857411.

Lacasta A., Morales-Hernández M., Murillo J., García-Navarro P. (2015) GPU implementation of the 2D shallow water equations for the simulation of rainfall/runoff events, Environmental Earth Sciences 74(11), 7295-7305. Doi: 10.1007/s12665-015-4215-z.

LeFloch P.G., Thanh M.D. (2007) The Riemann problem for the shallow water equations with discontinuous topography, Communications in Mathematical Sciences 5(4), 865-885. https://projecteuclid.org/euclid.cms/1199377555.

LeVeque R.J. (1992) Numerical methods for conservation laws, Birkhäuser, Basel.

LeVeque R.J., George D.L., Berger M.J. (2011) Tsunami modelling with adaptively refined finite volume methods, Acta Numerica 20, 2011-289. Doi: 10.1017/S0962492911000043.

Liang Q., Du G., Hall J.W., Borthwick A.G.L. (2008) Flood inundation modeling with an adaptive Quadtree Grid Shallow water Equation solver, ASCE Journal of Hydraulic Engineering 134(11), 1603-1610. Doi: 10.1061/(ASCE)0733-9429(2008)134:11(1603).

Martins R., Leandro J., Djordjević S. (2015) A well balanced Roe scheme for the local inertial equations with an unstructured mesh, Advances in Water Resources 83, 351-363. Doi: 10.1016/j.advwatres.2015.07.007.

Martins R., Leandro J., Djordjević S. (2016a) Analytical and numerical solutions of the Local Inertial Equations, International Journal of Non-Linear Mechanics 81, 222-229. Doi: 10.1016/j.ijnonlinmec.2016.01.015.

Martins R., Leandro J., Djordjević S. (2016b) Analytical solution of the classical dam-break problem for the Gravity Wave-Model equations, Journal of Hydraulic Engineering 142(5), 06016003. Doi: 10.1061/(ASCE)HY.1943-7900.0001121.

Martins R., Leandro J., Djordjević S. (2016c) Influence of sewer network models on urban flood damage assessment based on coupled 1D/2D models, Journal of Flood Risk Management (in press). Doi: 10.1111/jfr3.12244.

Martins R., Leandro J., Chen A.S., Djordjević S. (2017) A comparison of three dual drainage models: shallow water vs Local Inertial vs Diffusive Wave, Journal of Hydroinformatics 19(3), 331-348. Doi: 10.2166/hydro.2017.075.

Martins R., Leandro J., Chen A.S., Djordjević S. (2018) Wetting and drying numerical treatments for the Roe Riemann scheme, Journal of Hydraulic Research 56(2), 256-267. Doi: 10.1080/00221686.2017.1289256.

Mateo C.M.R., Yamazaki D., Kim H., Champathong A., Vaze J., Oki T. (2017) Impacts of spatial resolution and representation of flow connectivity on large-scale simulation of floods, Hydrology and Earth System Sciences 21(10), 5143-5163. Doi: 10.5194/hess-21-5143-2017.

Mignot E., Paquier A., Haider S. (2006) Modelling floods in a dense urban area using 2D shallow water equations, Journal of Hydrology 327(1-2), 186-199. Doi: 10.1016/j.jhydrol.2005.11.026.

Montuori C., Greco V. (1973) Fenomeni di moto vario a valle di una paratoia piana, L'Energia Elettrica 50(2),73–88 (in Italian).

Moramarco T., Melone F., Singh V.P. (2005) Assessment of flooding in urbanized ungauged basins: a case study in the Upper Tiber area, Italy, Hydrologic Processes 19(10), 1909-1924. Doi: 10.1002/hyp.5634.

Natale L., Savi F. (1991), Espansione di onde di sommersione su terreno inizialmente asciutto, Idrotecnica 1991(6), 397-406. (in Italian)

Neal J., Schumann G., Fewtrell T., Budimir M., Bates P., Mason D. (2011) Evaluating a new LISFLOOD-FP formulation with data from the summer 2007 floods in Tewkesbury, UK, Journal of Flood Risk Management 4(2), 88-95. Doi: 10.1111/j.1753-318X.2011.01093.x.

Neal J., Villanueva I., Wright N., Willis T., Fewtrell T., Bates P. (2012) How much physical complexity is needed to model flood inundation?, Hydrological Processes 26(15), 2264-2282. Doi: 10.1002/hyp.8339.

Nguyen N.Y., Ichikawa Y., Ishidaira H. (2016) Estimation of inundation depth using flood extent information and hydrodynamic simulations. Hydrological Research letters 10(1), 39-44. Doi: 10.3178/hrl.10.39.

Pistrika A., Tsakiris G., Nalbantis I. (2014) Flood depth-damage functions for built environment, Environmental Processes 1(4), 533-572. Doi: 10.1007/s40710-014-0038-2.

Pregnolato M., Ford A., Wilkinson S.M., Dawson R.J. (2017) The impact of flooding on road transport: A depth-disruption function, Transportation Research Part D: Transport and Environment 55, 67-81. Doi: 10.1016/j.trd.2017.06.020.

Price R.K. (1994) Flood Routing Models, In: Chaudhry M.H., Mays L.W. (eds) Computer Modeling of Free-Surface and Pressurized Flows, Springer, Dordrecht.

Savage J.T.S., Pianosi F., Bates P., Freer J., Wagener T. (2016) Quantifying the importance of spatial resolution and other factors through global sensitivity analysis of a flood inundation model, Water Resources Research 52(11), 9146-9163. Doi: 10.1002/2015WR018198.

Sielecki A. (1968) An energy-conserving difference scheme for the storm surge equations, Monthly Weather Review 96(3), 150-156. Doi: 10.1175/1520-0493(1968)096<0150:AECDSF>2.0.CO;2.

Sobey R.J. (2009) Wetting and drying in coastal flows, Coastal Engineering 56(5-6), 565-576. Doi: 10.1016/j.coastaleng.2008.12.001.

Thacker W.C. (1981) Some exact solutions to the nonlinear shallow-water wave equations, Journal of Fluid Mechanics 107, 499-508. Doi: 10.1017/S0022112081001882.

Toro E.F. (2001) Shock-capturing methods for free-surface flows, Wiley, Chichester.

Toro E.F. (2009) Riemann solvers and numerical methods for Fluids Dynamics, A practical introduction, Springer-Verlag, Berlin.

Toro E.F., Garcia-Navarro P. (2007) Godunov-type methods for free-surface shallow flows: A review, IAHR Journal of Hydraulic Research 45(6), 736-751. Doi: 10.1080/00221686.2007.9521812.

Tsai C.W. (2003) Applicability of kinematic, noninertia, and quasi-steady dynamic wave models to unsteady flow routing, Journal of Hydraulic Engineering 129(8), 613-627. Doi: 10.1061/(ASCE)0733-9429(2003)129:8(613).

Uusitalo S. (1960) The numerical calculation of wind effect on sea level elevations, Tellus, 1960, 12(4), 427-435. Doi: 10.1111/j.2153-3490.1960.tb01329.x.

Wang Y., Liang Q., Kesserwani G., Hall J.W. (2011) A 2D shallow flow model for practical dam-break simulations, IAHR Journal of Hydraulic Research 49(3), 307-316. Doi: 10.1080/00221686.2011.566248.

Whitham G.B. (1955) The effects of hydraulic resistance in the dam-break problem, Proocedings of the Royal Society A 227(1170), 399-407. Doi: 10.1098/rspa.1955.0019.

Whitham G.B. (1974) Linear and nonlinear waves. Wiley Interscience, New York.

Xing Y., Shu C.-W. (2005) High order finite difference WENO schemes with the exact conservation property for the shallow water equations, Journal of Computational Physics 2018(1), 206-227. doi: 10.1016/j.jcp.2005.02.006.

Yamazaki D., de Almeida G.A.M., Bates P. (2013) Improving computational efficiency in global river models by implementing the local inertial flow equation and a vector-based river network map, Water Resources Research 49(11), 7221-7235. Doi: 10.1002/wrcr.20552.

Yamazaki D., Tanaka T., Bates P. (2015) Rapid and stable flood inundation modelling using the Local Inertial Equation, Journal of the Japan Society of Hydrology and Water Resources 28(3), 124-130. Doi: 10.3178/jjshwr.28.124. (in Japanese)

**Figures**

Figure 1. Tip-region of the wave propagating on a sloping floodplain with friction: comparison between the SWE model (continuous line) and LInA model (dashed line).

Figure 2. Riemann problem on horizontal bed. Graphical solutions for the initial data in Table 1.

Figure 3. Riemann problem on horizontal dry bed. Exact solution at $t = 5$ s (flow depth). Rarefaction-shock with $h_L = 1$ m, $h_L u_L = 0.20$ m$^2$/s (a); shock-shock with $h_L = 1$ m, $h_L u_L = 5$ m$^2$/s (b).

Figure 4. Riemann problem on horizontal dry bed. Comparison of the wave speeds.

Figure 5. Impact on a wall. Exact solution a $t = 5$ s (flow depth) for $h_L = 1$ m and $h_L u_L = 1.5$ m$^2$/s (a); ratio $h_{LInA}/h_{SWE}$ at the wall for different values of $F(\mathbf{u}_L)$ (b).

Figure 6. Wetting with SWE and LInA of the horizontal floodplain with friction.

Figure 7. Riemann problem at the dry bed step. Exact solution at $t = 5$ s (free surface elevation) for the initial data in Table 2.

Figure 8. Riemann problem at the dry bed drop. Exact solution at $t = 5$ s (free surface elevation) for the initial data in Table 3.

Figure 9. Limits of existence for the Riemann problem at the bed drop.

Figure 10. Finite-difference q-centred scheme by de Almeida et al. (2012). Flow depth at time t = 5.002 s for Test 1 (a), and Test 2 (b).

Figure 11. Finite-difference q-centred scheme by de Almeida et al. (2012). Free surface elevation for the Test 3 at times $t_1 = 350.17$ s (a), $t_2 = 710.39$ s (b), $t_3 = 1060.57$ s (c), and $t_4 = 1420.73$ s (d).

Figure 12. Modified finite-difference q-centred scheme. Flow depth at time t = 5.001 s for Test 1 (a), and at time t = 5.002 s for Test 2 (b).

Figure 13. Modified finite-difference q-centred scheme. Free surface elevation for the Test 3 at times $t_1 = 350.24$ s (a), $t_2 = 710.52$ s (b), $t_3 = 1060.79$ s (c), and $t_4 = 1421.02$ s (d).

Figure 14. Rusanov Finite Volume scheme. Flow depth at time t = 5.002 s for Test 1 (a), and Test 2 (b).

Figure 15. Rusanov Finite Volume scheme. Free surface elevation for the Test 3 at times $t_1 = 350.17$ s (a), $t_2 = 710.39$ s (b), $t_3 = 1060.57$ s (c), and $t_4 = 1420.73$ s (d).

Figure 16. Modified Rusanov Finite Volume scheme. Flow depth at time t = 5.002 s for Test 1 (a), and at time t = 5.001 s for Test 2 (b).

Figure 17. Modified Rusanov Finite Volume scheme. Free surface elevation for the Test 3 at times $t_1 = 350.22$ s (a), and $t_2 = 710.41$ s (b).

**Tables**

Table 1. Riemann problem on horizontal bed. Initial data for the example solutions in Figure 2.

Table 2. Riemann problem at the dry bed step. Initial data for the example solutions in Figure 7.

Table 3. Riemann problem at the dry bed drop. Initial data for the example solutions in Figure 8.

**Table 1. Riemann problem on horizontal bed. Initial data for the example solutions in Figure 2.**

| Configuration | $h_L$ (m) | $u_L$ (m/s) | $h_R$ (m) | $u_R$ (m/s) |
|---|---|---|---|---|
| No solution (Fig. 1a) | 0.50 | 1.00 | 0.50 | 5 |
| Rarefaction – rarefaction (Fig. 1b) | 1 | 1 | 1 | 3 |
| Shock – rarefaction (Fig. 1c) | 1 | 3 | 2 | 1.5 |
| Shock – shock (Fig. 1d) | 1 | 3 | 1 | 1 |

**Table 2. Riemann problem at the dry bed step. Initial data for the example solutions in Figure 7.**

| Example solution | $h_L$ (m) | $u_L$ (m/s) | $z_L$ (m/s) | $h_R$ (m) | $u_R$ (m/s) | $z_R$ (m/s) |
|---|---|---|---|---|---|---|
| Figure 5a | 1.00 | 0.20 | 0 | 0 | 0 | 0.50 |
| Figure 5b | 1.00 | 1.50 | 0 | 0 | 0 | 1.45 |
| Figure 5c | 1.00 | 5.00 | 0 | 0 | 0 | 0.50 |
| Figure 5d | 1.00 | 5.00 | 0 | 0 | 0 | 0.10 |

**Table 3. Riemann problem at the dry bed drop. Initial data for the example solutions in Figure 8.**

| Example solution | $h_L$ (m) | $u_L$ (m/s) | $z_L$ (m/s) | $h_R$ (m) | $u_R$ (m/s) | $z_R$ (m/s) |
|---|---|---|---|---|---|---|
| Figure 6a | 1.00 | 0.20 | 0 | 0 | 0 | -0.20 |
| Figure 6b | 1.00 | 0.20 | 0 | 0 | 0 | -1.01 |
| Figure 6c | 0.18 | 1.11 | 0 | 0 | 0 | -0.01 |
| Figure 6d | 1.00 | 5.00 | 0 | 0 | 0 | -1 |

Figure 1. Tip-region of the wave propagating on a sloping floodplain with friction: comparison between the SWE model (continuous line) and LInA model (dashed line).

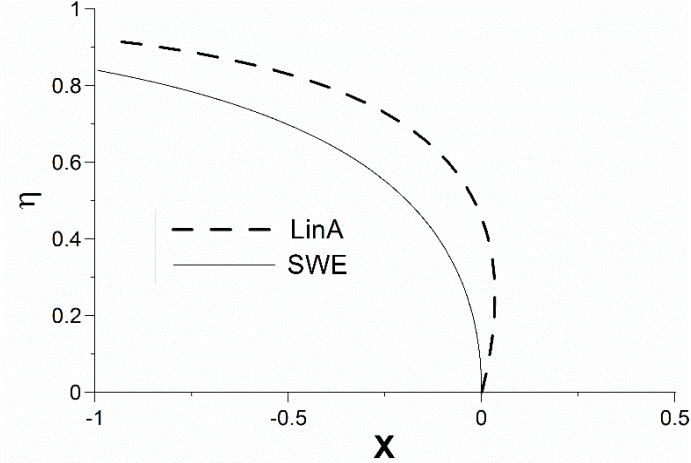

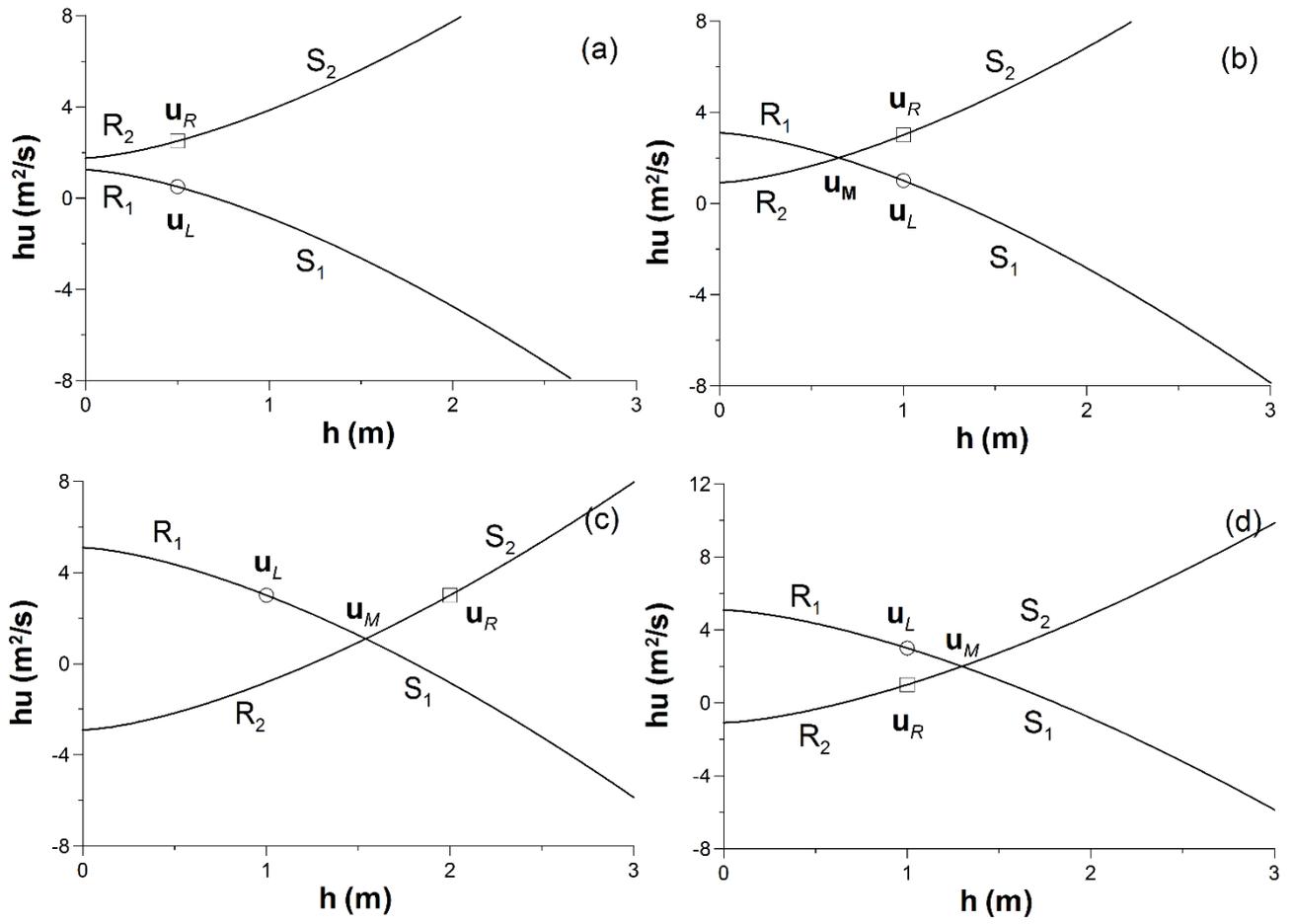

Figure 2. Riemann problem on horizontal bed. Graphical solutions for the initial data in Table 1.

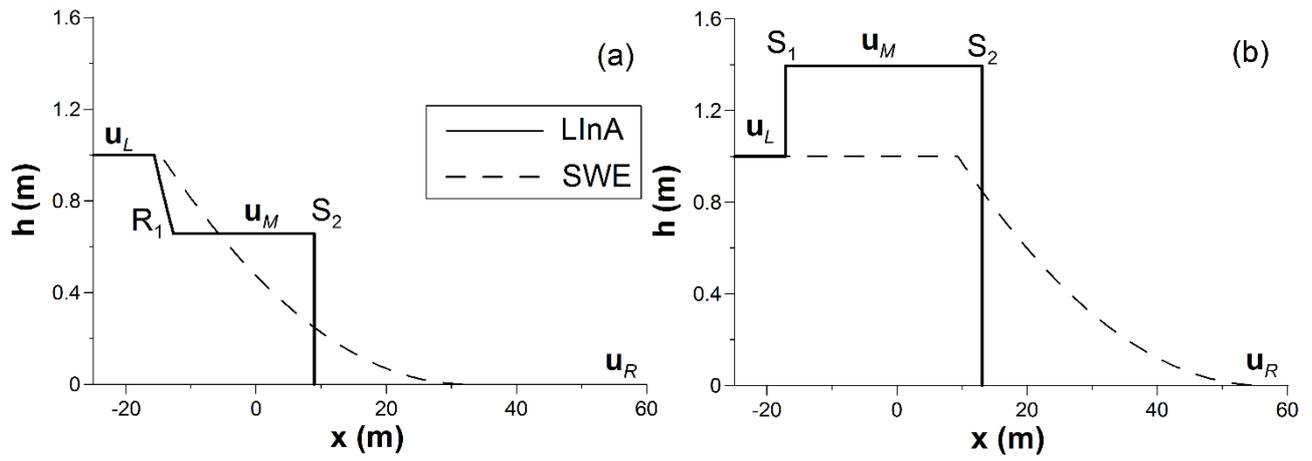

Figure 3. Riemann problem on horizontal dry bed. Exact solution at $t = 5$ s (flow depth). Rarefaction-shock with $h_L = 1$ m, $h_L u_L = 0.20$ m$^2$/s (a); shock-shock with $h_L = 1$ m, $h_L u_L = 5$ m$^2$/s (b).

Figure 4. Riemann problem on horizontal dry bed. Comparison of the wave speeds.

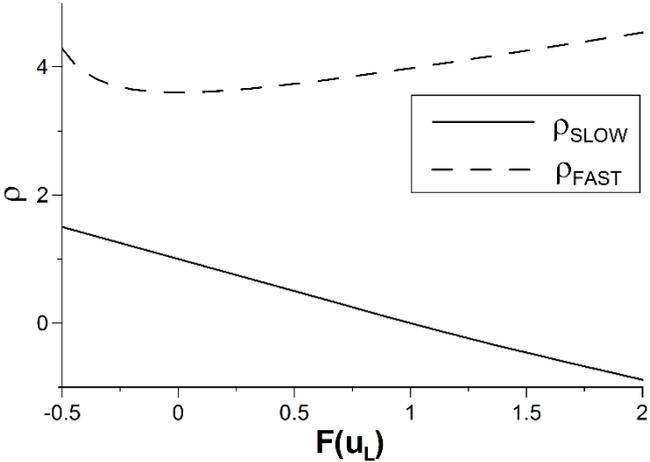

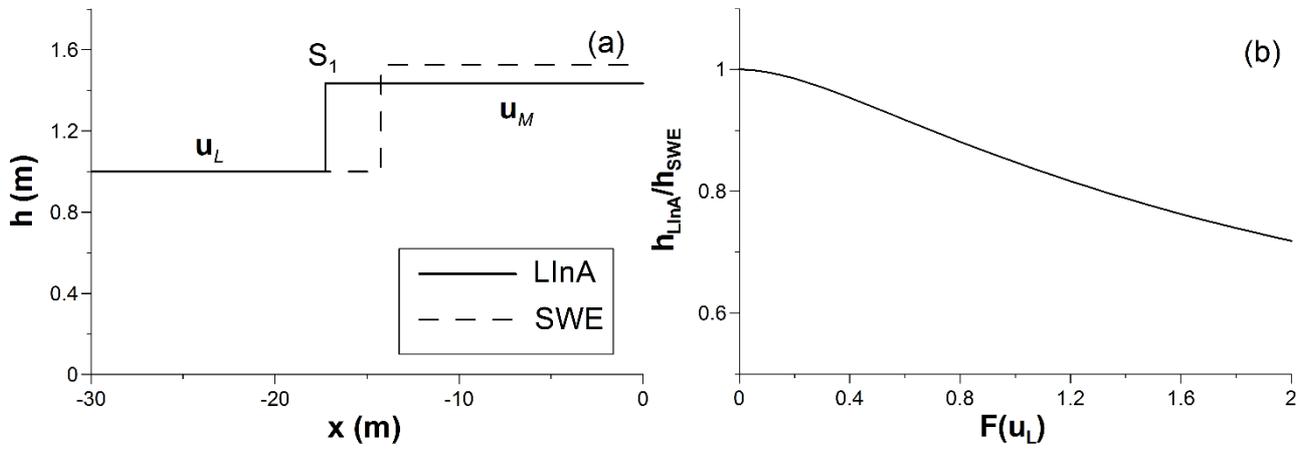

Figure 5. Impact on a wall. Exact solution a $t = 5$ s (flow depth) for $h_L = 1$ m and $h_L u_L = 1.5$ m$^2$/s (a); ratio $h_{LInA}/h_{SWE}$ at the wall for different values of $F(u_L)$ (b).

Figure 6. Wetting with SWE and LInA of the horizontal floodplain with friction.

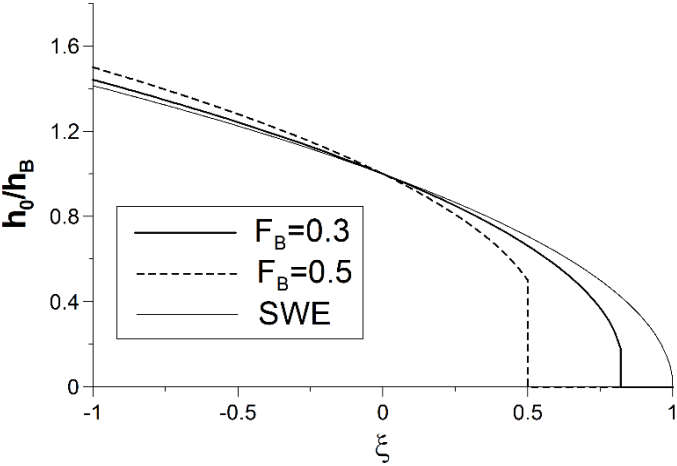

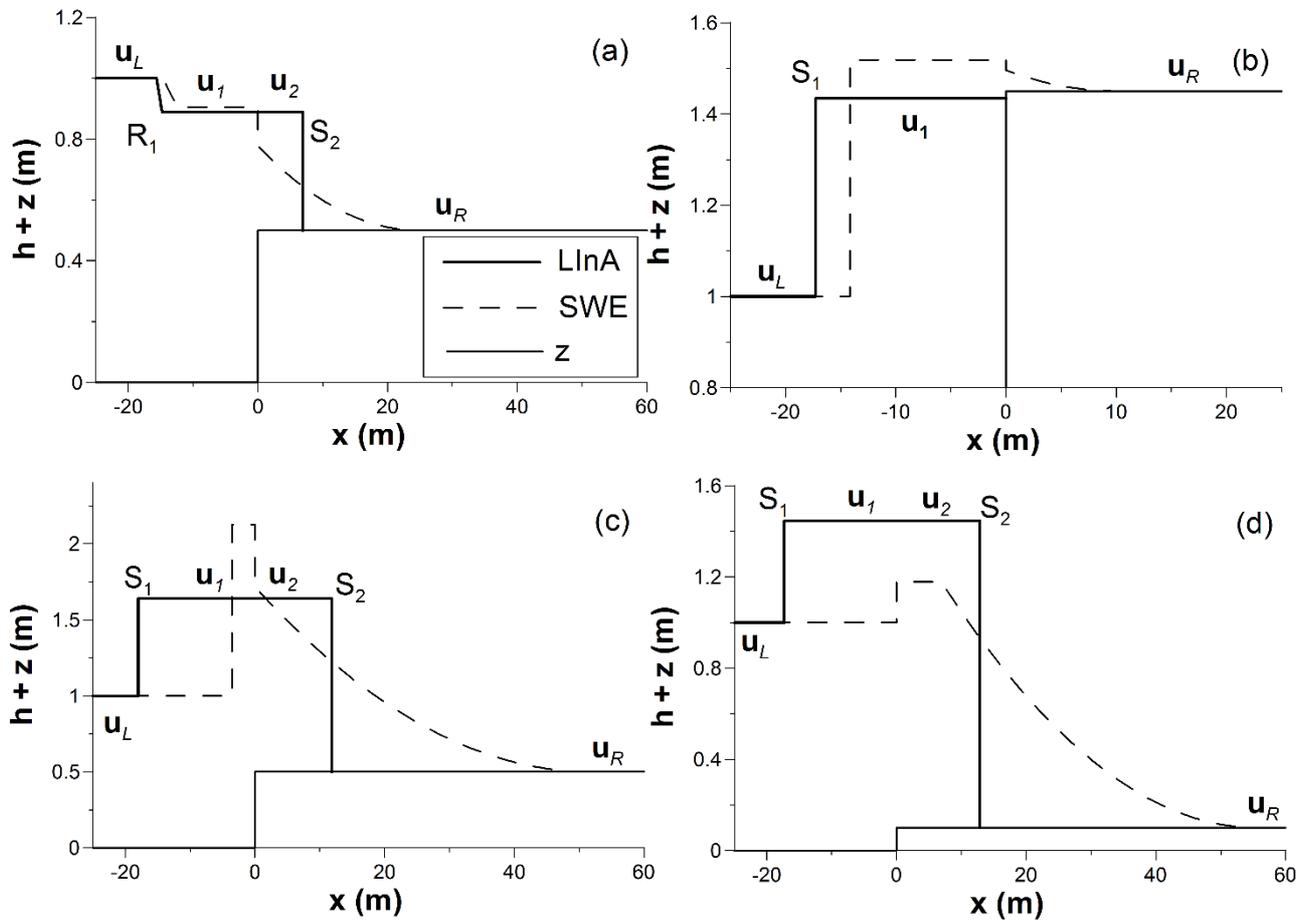

Figure 7. Riemann problem at the dry bed step. Exact solution at $t = 5$ s (free surface elevation) for the initial data in Table 2.

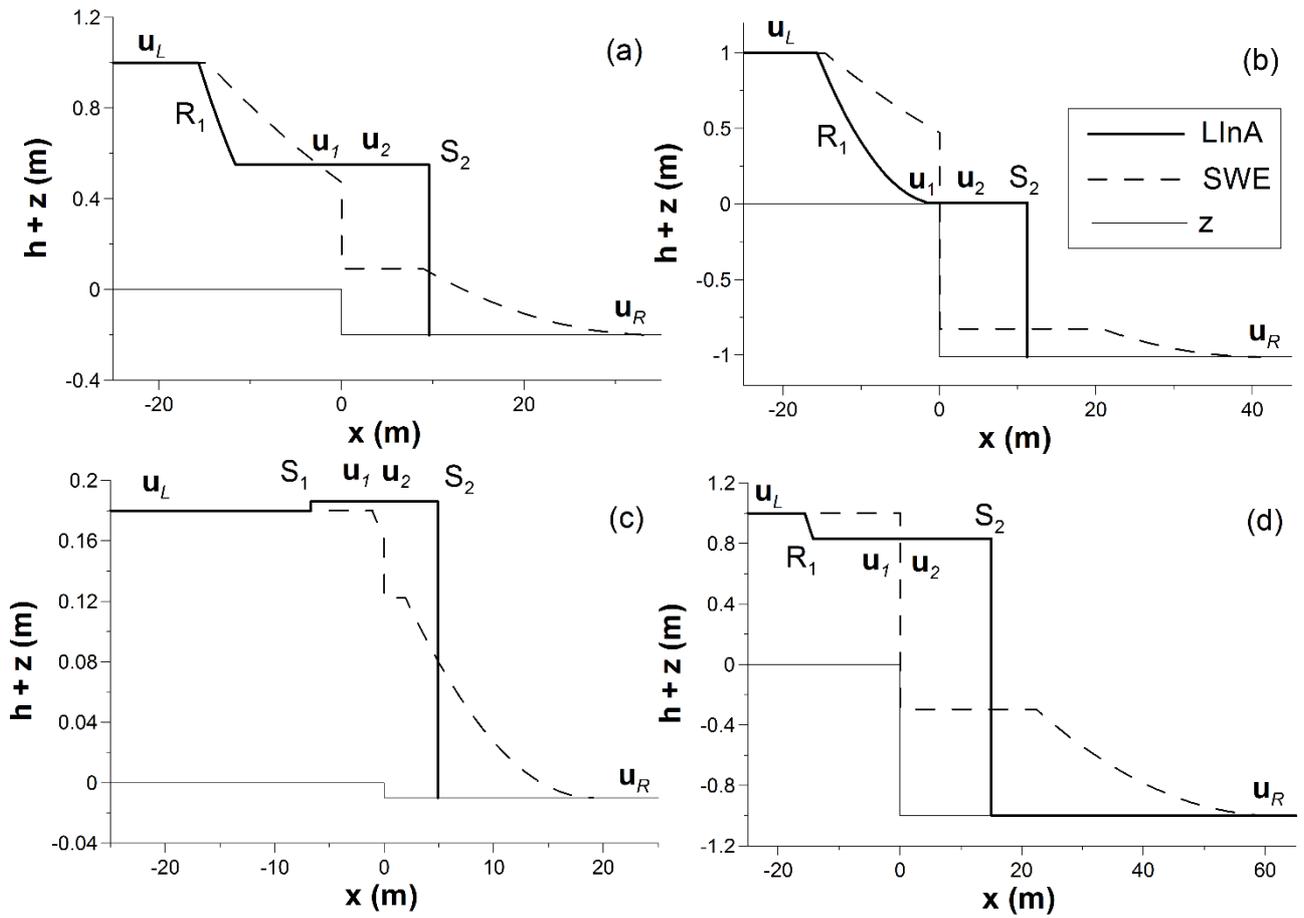

Figure 8. Riemann problem at the dry bed drop. Exact solution at $t = 5$ s (free surface elevation) for the initial data in Table 3.

Figure 9. Limits of existence for the Riemann problem at the bed drop.

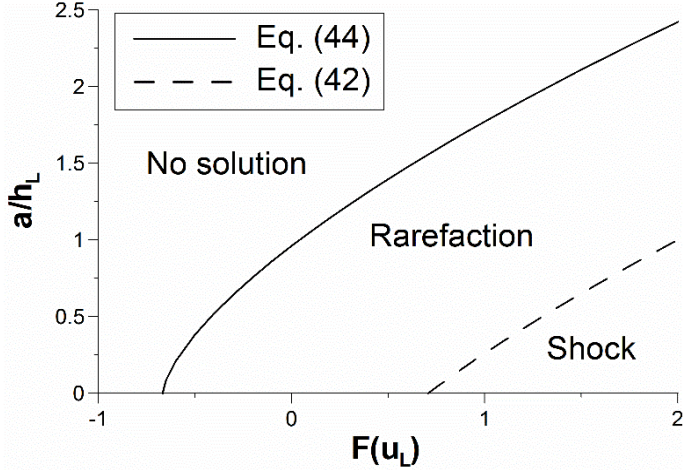

Figure 10. Finite-difference q-centred scheme by de Almeida et al. (2012). Flow depth at time t = 5.002 s for Test 1 (a), and Test 2 (b).

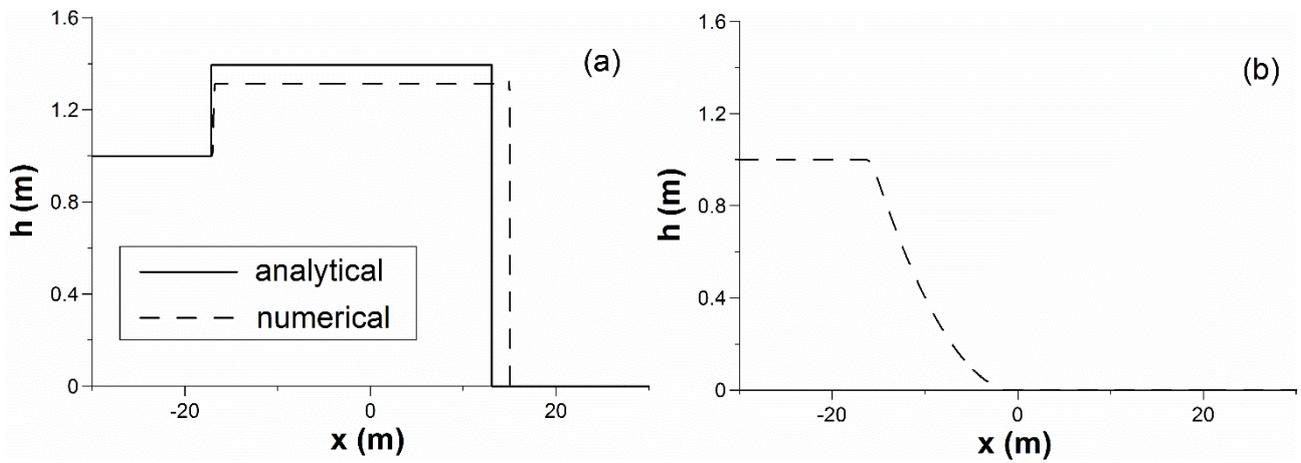

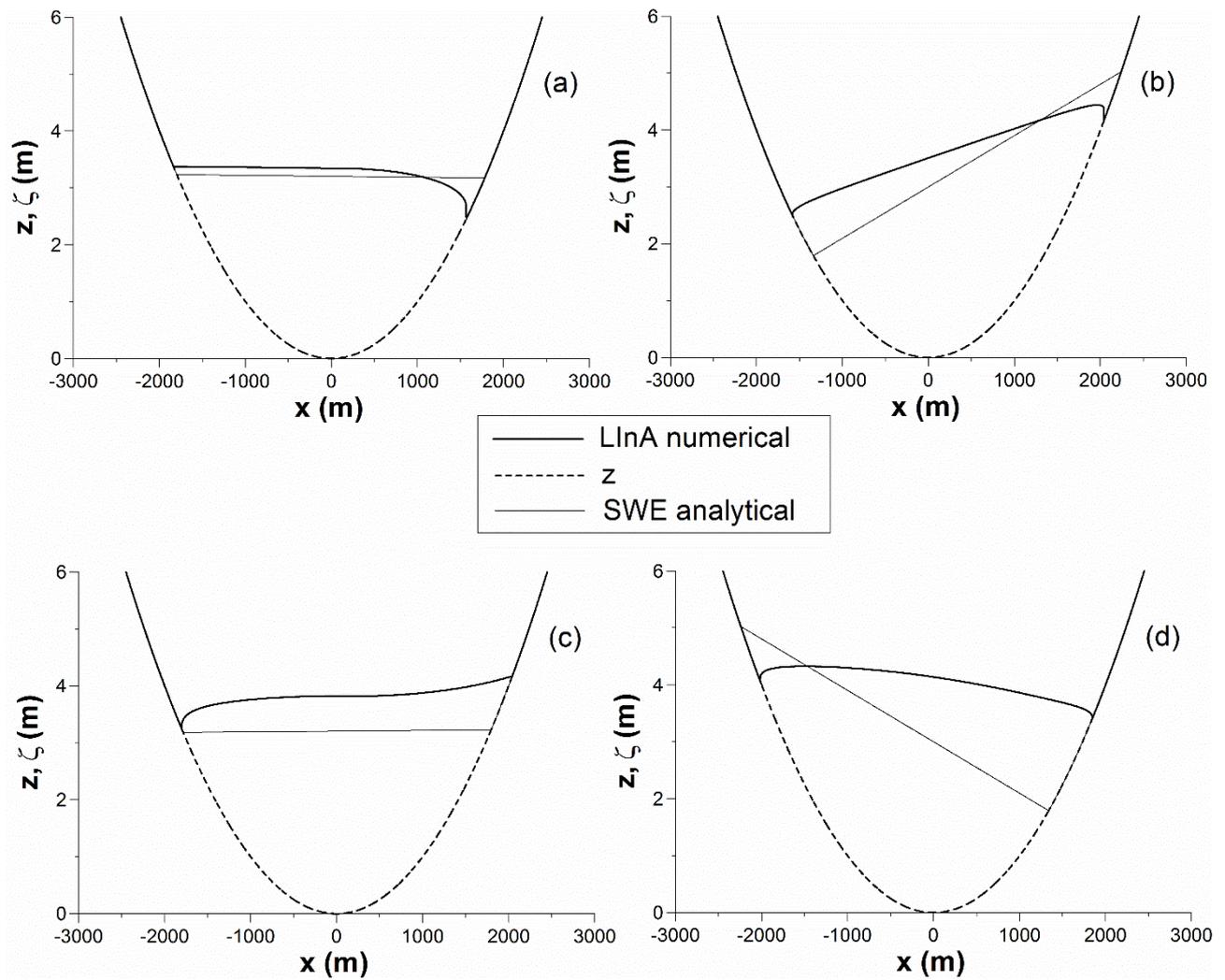

Figure 11. Finite-difference q-centred scheme by de Almeida et al. (2012). Free surface elevation for the Test 3 at times $t_1 = 350.17$ s (a), $t_2 = 710.39$ s (b), $t_3 = 1060.57$ s (c), and $t_4 = 1420.73$ s (d).

Figure 12. Modified finite-difference q-centred scheme. Flow depth at time t = 5.001 s for Test 1 (a), and at time t = 5.002 s for Test 2 (b).

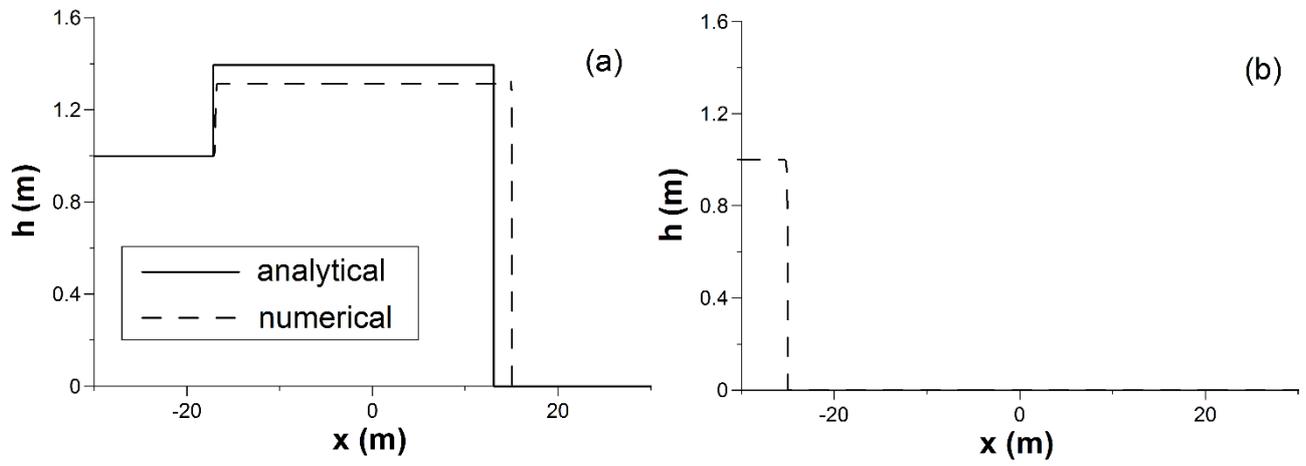

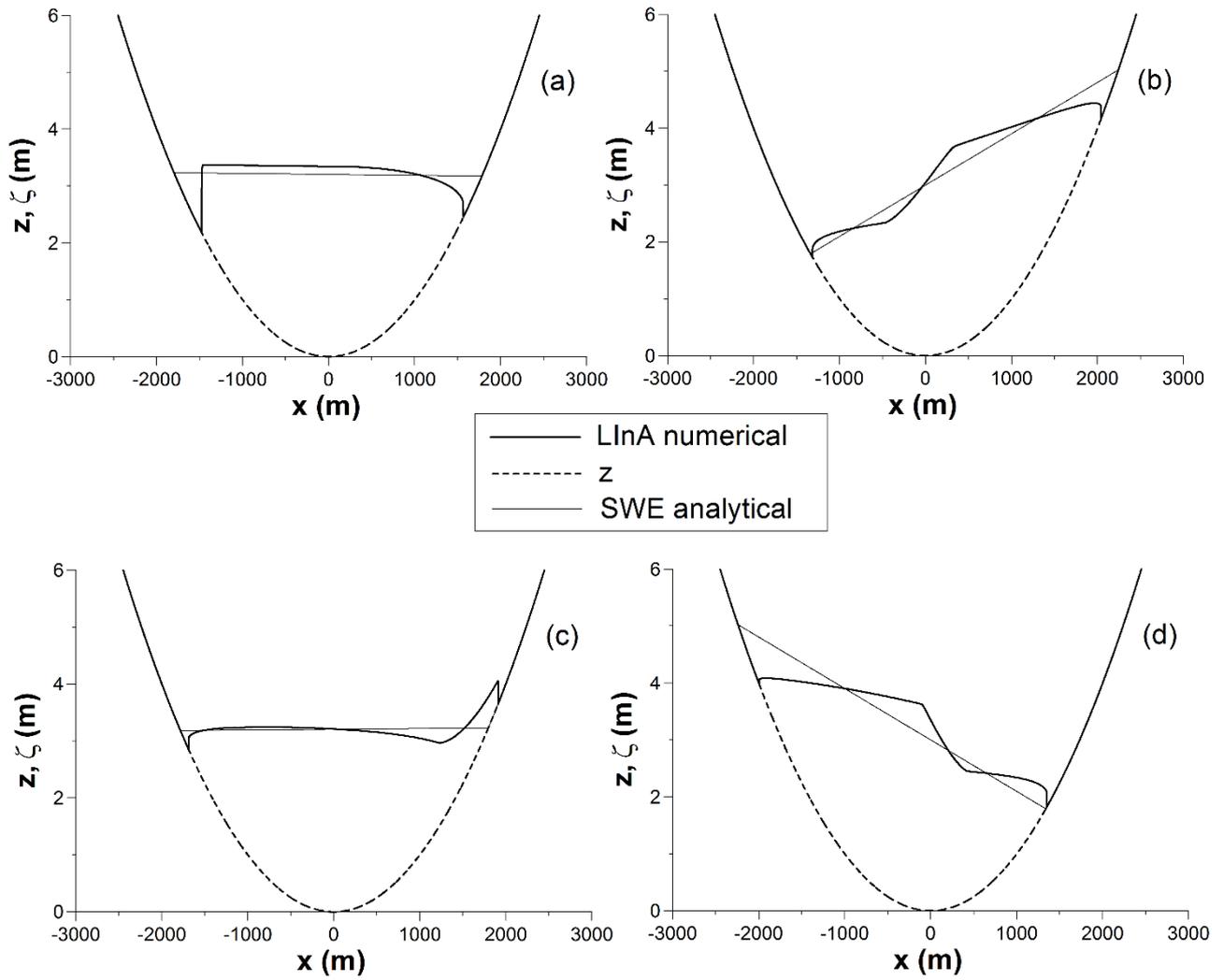

Figure 13. Modified finite-difference q-centred scheme. Free surface elevation for the Test 3 at times $t_1 = 350.24$ s (a), $t_2 = 710.52$ s (b), $t_3 = 1060.79$ s (c), and $t_4 = 1421.02$ s (d).

Figure 14. Rusanov Finite Volume scheme. Flow depth at time t = 5.002 s for Test 1 (a), and Test 2 (b).

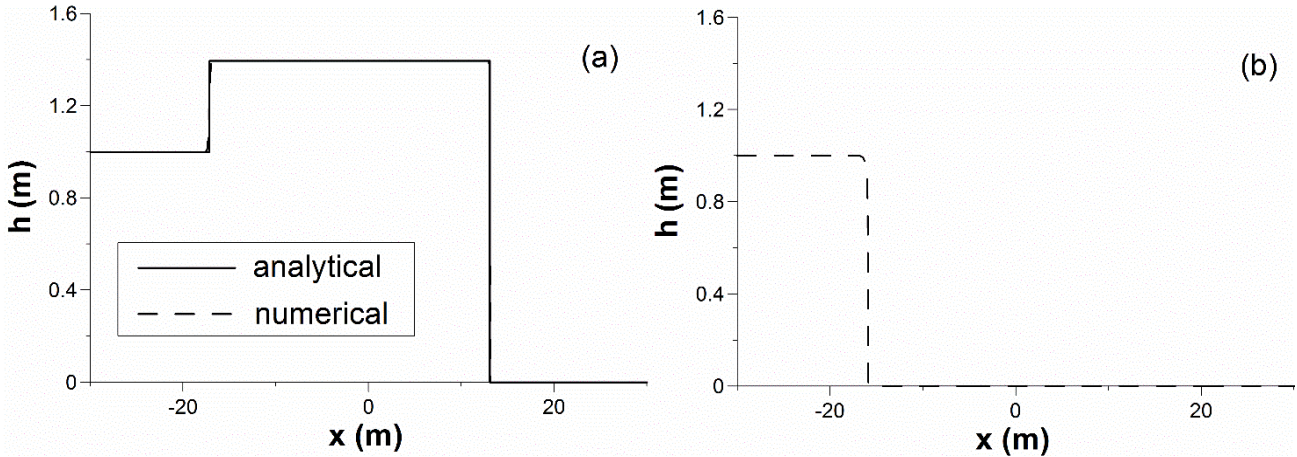

Figure 15. Rusanov Finite Volume scheme. Free surface elevation for the Test 3 at times $t_1$ = 350.17 s (a), $t_2$ = 710.39 s (b), $t_3$ = 1060.57 s (c), and $t_4$ = 1420.73 s (d).

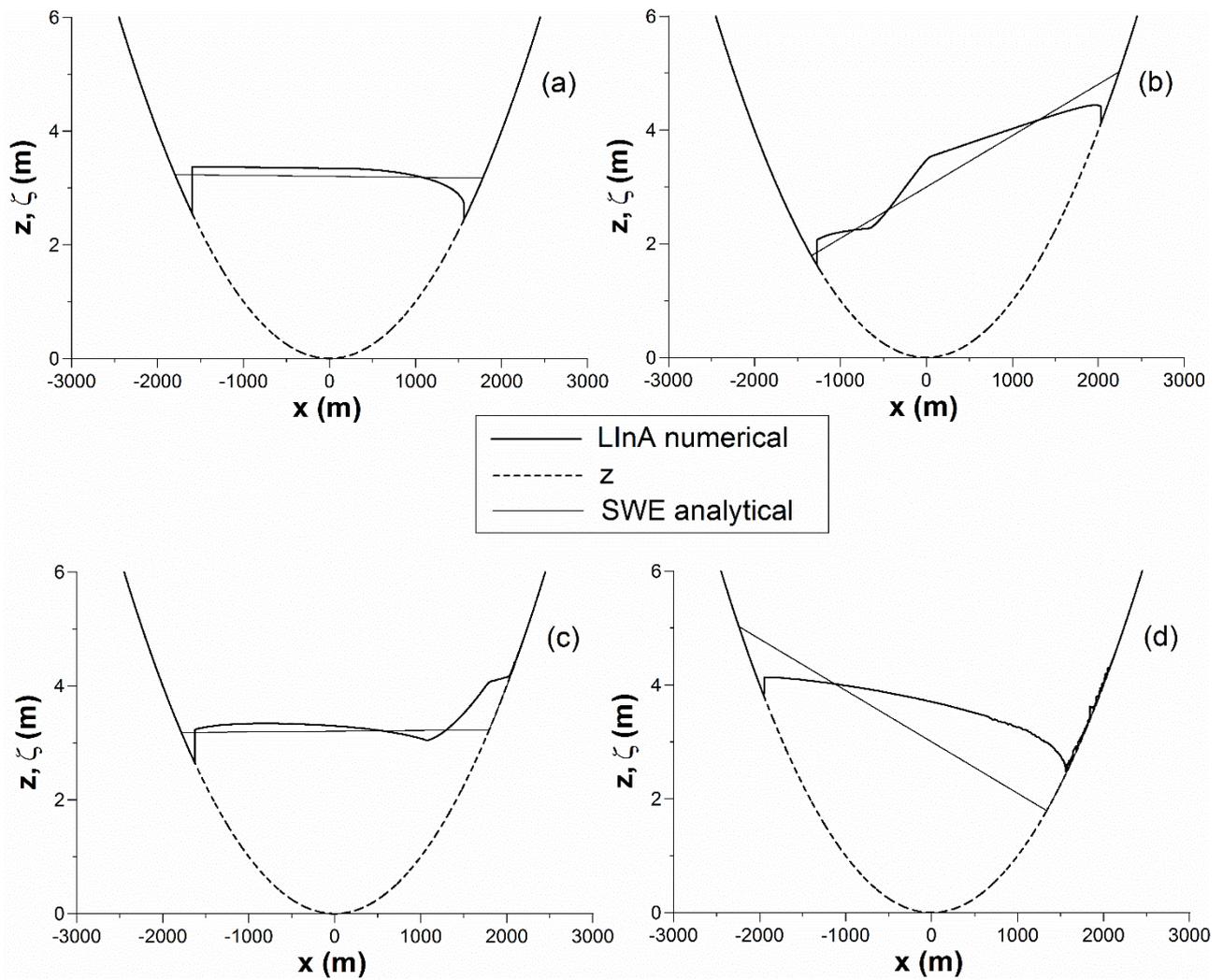

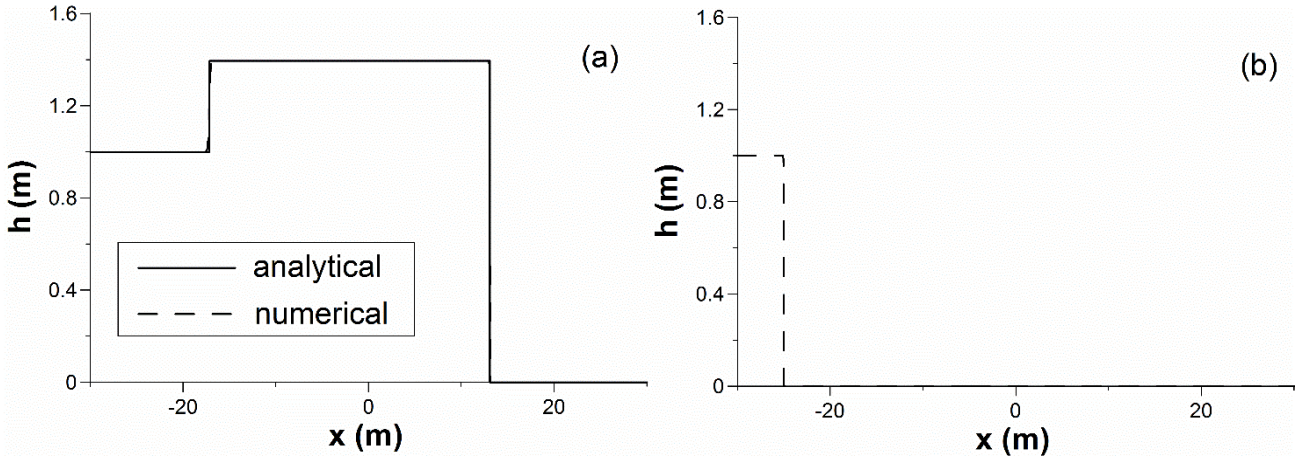

Figure 16. Modified Rusanov Finite Volume scheme. Flow depth at time t = 5.002 s for Test 1 (a), and at time t = 5.001 s for Test 2 (b).

Figure 17. Modified Rusanov Finite Volume scheme. Free surface elevation for the Test 3 at times $t_1$ = 350.22 s (a), and $t_2$ = 710.41 s (b).

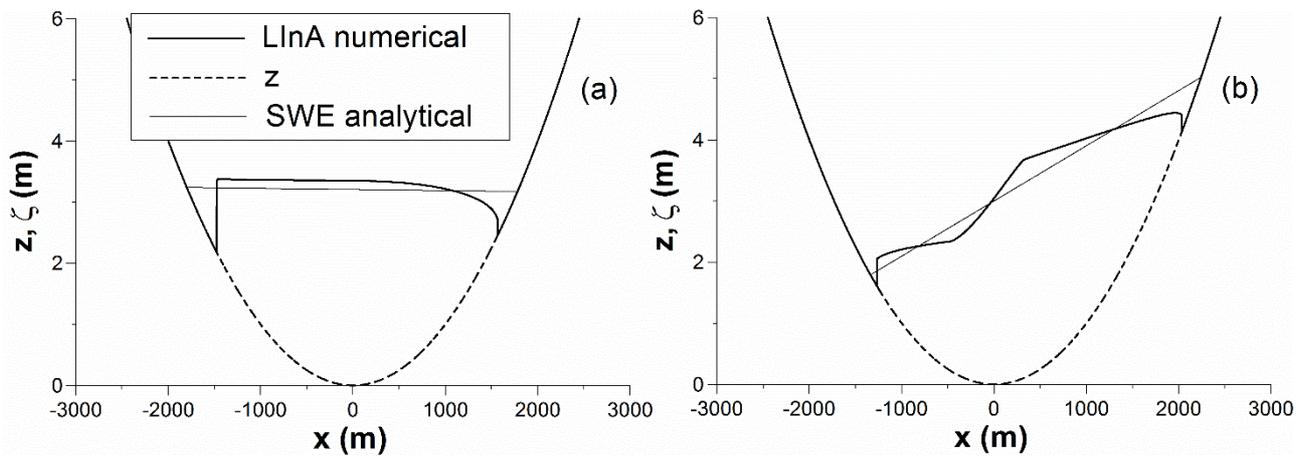